\documentclass[onecolumn,notitlepage,secnumarabic,amssymb,nobibnotes,aps,pre,superscriptaddress]{revtex4-1}

\usepackage{graphicx}
\usepackage{amsmath}
\usepackage[dvipsnames]{xcolor}

\newcommand{\nn}{\nonumber}
\newcommand{\eq}[1]{Eq.~\eqref{#1}}

\newcommand{\fig}[1]{Fig.~\ref{#1}}
\newcommand{\sctn}[1]{\S~\ref{#1}}

\newcommand{\movie}[1]{Movie~#1}
\newcommand{\movies}[1]{Movies~#1}
\newcommand{\ie}{\textit{i.e.} }

\newcommand{\abs}[1]{\left| #1 \right|}
\renewcommand{\vec}[1]{\underline{#1}}
\newcommand{\tens}[1]{\underline{\underline{#1}}}
\newcommand{\del}{\vec{\nabla}}

\newcommand{\RE}[1]{\text{Re}\left[#1\right]}

\newcommand{\tr}[1]{\mathrm{tr}\left( #1 \right) }
\newcommand{\diag}[1]{\mathrm{diag}\left( #1 \right) }
\renewcommand{\det}[1]{\mathrm{det}\left( #1 \right) }
\newcommand{\commutator}[2]{\left[ #1 , #2 \right]  } 

\newcommand{\ps}{\abs{\psi}}
\newcommand{\app}{\ps^2}
\newcommand{\thick}{d_0}

\begin{document}

\title[Smectic Layering]{Smectic Layering:\\
 Landau theory for a complex-tensor order parameter}

\author{Jack Paget}
\address{Interdisciplinary Centre for Mathematical Modelling and Department of Mathematical Sciences, Loughborough University, Loughborough, Leicestershire LE11 3TU, UK.}
\author{Una Alberti}
\affiliation{School of Physics and Astronomy, The University of Edinburgh, Peter Guthrie Tait Road, Edinburgh, EH9 3FD, UK.}
\author{Marco G. Mazza}
\affiliation{Interdisciplinary Centre for Mathematical Modelling and Department of Mathematical Sciences, Loughborough University, Loughborough, Leicestershire LE11 3TU, UK.}
\affiliation{Max Planck Institute for Dynamics and Self-Organization (MPIDS), Am Fa{\ss}berg 17, D-37077 G\"{o}ttingen, Germany.}
\author{Andrew J. Archer}
\affiliation{Interdisciplinary Centre for Mathematical Modelling and Department of Mathematical Sciences, Loughborough University, Loughborough, Leicestershire LE11 3TU, UK.}
\author{Tyler N. Shendruk}
\affiliation{School of Physics and Astronomy, The University of Edinburgh, Peter Guthrie Tait Road, Edinburgh, EH9 3FD, UK.}
\email{t.shendruk@ed.ac.uk}

\begin{abstract}
Composed of microscopic layers that stack along one direction while maintaining fluid-like positional disorder within layers, smectics are excellent systems for exploring topology, defects and geometric memory in complex confining geometries. 
However, the coexistence of crystalline-like characteristics in one direction and fluid-like disorder within layers makes lamellar liquid crystals notoriously difficult to model---especially in the presence of defects and large distortions.
Nematic properties of smectics can be comprehensively described by the $\tens{Q}$-tensor but to capture the features of the smectic layering alone, we develop a phenomenological Landau theory for a complex-tensor order parameter $\tens{E}$, which is capable of describing the local degree of lamellar ordering, layer displacement, and orientation of the layers. 
This theory can account for both parallel and perpendicular elastic contributions. 
In addition to resolving the potential ambiguities inherent to complex scalar order parameter models, this model reduces to previous employed models of simple smectics, and opens new possibilities for numerical studies on smectics possessing many defects, within complex geometries and under extreme confinement.
\end{abstract}

\maketitle

%%-------------------------------------------------------------------%%
%%-------------------------------------------------------------------%%
%%-------------------------------------------------------------------%%
%% Intro %%
%%-------------------------------------------------------------------%%
%%-------------------------------------------------------------------%%
%%-------------------------------------------------------------------%%
\section{Introduction}\label{sctn:intro}

Many liquid crystaline mesophases exist between isotropic liquids and crystaline solids, including the family of materials known as {\em smectic liquid crystals}. 
Smectics are notable because smectogen molecules tend to reside in layers, gaining a limited degree of spatial ordering. 
As essentially stacks of 2D fluids, simple smectics are {\em lamellar}---layered but possessing little-to-no positional order within layers--- with broken translational symmetry in only one spatial direction. 
While the shape of the lamellae need not be flat, orientation of the layers necessarily breaks rotational symmetry---whereby flipping the local layer normal direction is physically inconsequential. 
They share, and often directly inherit, this orientational symmetry from {\em nematic} liquid crystals in which rod-like molecules tend to align without any positional ordering. 
The typical, but not universal~\cite{Mukherjee2021b}, sequence of phases with decreasing temperature is isotropic fluid $\to$ nematic $\to$ smectic $\to$ crystalline solid. 

Beyond the simple definition of smectics as ordered monolayers of fluids, smectics are a diverse group of mesophases and there exists a multitude of definitions specifying molecular orientation or positioning within the layers~\cite{Baron2001}. 
In terms of orientation, when the smectogen molecules tend to align parallel to the local layer normal axis the smectic is denoted {\em sm-A}~\cite{degennes1972, mcmillan1971,chen1976,Meyer2021}. 
On the other hand, {\em sm-C} are mesophases in which the smectogens are tilted within the layers~\cite{McMillan1973,chen1976,Lagerwall2006,Mukherjee2021a}. 
Beyond these relatively simple cases lies a zoo of more complicated smectic phases that possess some degree of positional ordering within the layers, such as hexatic {\em sm-B} phases~\cite{prost1984}. 
However, in all these cases, smectics possess long-range positional order in only a single direction and fluid layers can viscously slide over each other. 
The layer thickness varies with temperature and microscopic details, but is typically on the scale of the molecular length. 

% The division between {\em thermotropic} and {\em lyotopic} materials is particularly important when considering smectics. 
% Lyotropics are commonly composed of amphiphilic molecules in water and it is the relative concentration that governs the phase. 
% Many structures are possible including lamellar phases, in which bilayers of amphiles form a smectic. 
% In such materials, deformations that distort the orientation of the layers are far easier than compression or dilation of the layer thickness. 
% Thermotropics are composed of rod-like mesogens, whose anisotropic shape dictates the character of the liquid crystal phase as a function of temperature. 
% In these materials, the elasticity associated with the response distortions, such as splay (or twist and bend), can be significant. 
% The distinct between thermotropics and lyotropics is not perfect and many materials exhibit combined properties. 

As materials possessing a broken rotational symmetry and partial translational symmetry breaking, smectics are excellent materials for exploring questions of topology and self-assembly in soft condensed matter physics. 
Colloidal inclusions in smectics exhibit fascinating defect structures in the vicinity of the colloid~\cite{harth2009, gharbi2018, rasi2018, Dolganov2019} and, conversely, defect patterns can be used to guide colloidal assembly~\cite{Honglawan2015, do2020}. 
By microfabricating surfaces that confine the fluid, complex smectic conformation can be achieved~\cite{preusse2020, kim2009a, kim2009b, kim2010a, kim2010b}. 
Similarly, confining smectic in porous environments can create steady states of defects, which govern the ability of the fluid to flow~\cite{Bandyopadhyay2005}. 
Extreme confinements can be achieved using colloidal liquid crystals~\cite{Cortes2016, wittmann2021} or by freely-suspending films~\cite{cluzeau2001, Nguyen2010, radzihovsky2017, selimi2017}. 

However, these same properties contrive to make lamellar fluids notoriously difficult to model and understanding their structural dynamics in complex geometries remains an outstanding challenge. 
Molecular dynamics~\cite{frenkel1988,Schulz2014,zhang2021}, Monte-Carlo~\cite{puschel2017,Monderkamp2021,monderkamp2022}, density functional approaches~\cite{vitral2019,Vitral2021,wittmann2021} and neural networks~\cite{Schneider2021} have offered exciting results, but a major hurdle remains the difficulty of performing numerical simulations on macroscopic length scales. 
The primary source of difficulty faced by macroscopic theories is that complex scalar order parameters for smectics are not truly a single-valued functions of position and so possess ambiguity~\cite{Pevnyi2014,chen2009}. 
This is not a fatal issue for theoreticians who can work locally in the vicinity of large distortions, such as defect singularities~\cite{ambrozic2004,alexander2012}. 
However, it causes crippling issues for numerical approaches to macroscopic theories in the vicinity of defects, which require an order parameter that is single-valued on a global coordinate system that has been discretized on a computational mesh.  
The situation is analogous to the state of nematics, in which branch cuts in the vicinity of defects are required when the system is described by a director in Frank--Oseen and Ericksen--Leslie but are avoided by $\tens{Q}$-tensor theory. 

In a recent attempt to address this short-coming, we proposed modelling simple smectic using a complex, symmetric, traceless, uniaxial, normal, globally gauge invariant tensorial smectic order parameter $\tens{E}$~\cite{paget2022}. 
We demonstrated an $\tens{E}$-based formalism is capable of describing both disclination and dislocation type defects, glassy/disordered configurations and smectic ordering around colloidal inclusions. 
Here, we extend this work to present a more detailed $\tens{E}$-theory. 
In particular, we begin by considering the background of lamellar systems (\sctn{sctn:layers}) and $\tens{Q}$-tensor theory for nematics (\sctn{sctn:Q}) before presenting our argument for the form of $\tens{E}$ (\sctn{sctn:E}). 
We then construct a phenomenological Landau theory that employs $\tens{E}$ in the bulk, compression and curvature free energies (\sctn{sctn:freeen}), which is analogous to the Landau-de~Gennes theory for nematics~\cite{Andrienko2018}. 
By considering the constraints discussed in \sctn{sctn:E}, we present a numerical evolution scheme to preserve the form of $\tens{E}$ (\sctn{sctn:lagrange}).
We gain physical insight into these terms through non-dimensionalization (\sctn{sctn:nondim}) and by applying simplifying assumptions, which demonstrate that $\tens{E}$-theory reduces to simple models of smectic distortion (\sctn{sctn:reducing}).

%%-------------------------------------------------------------------%%
%%-------------------------------------------------------------------%%
%%-------------------------------------------------------------------%%
%% Layers %%
%%-------------------------------------------------------------------%%
%%-------------------------------------------------------------------%%
%%-------------------------------------------------------------------%%
\subsection{Lamellae}\label{sctn:layers}
\begin{figure*}[tb]
    \centering
    \includegraphics[width=0.9\textwidth]{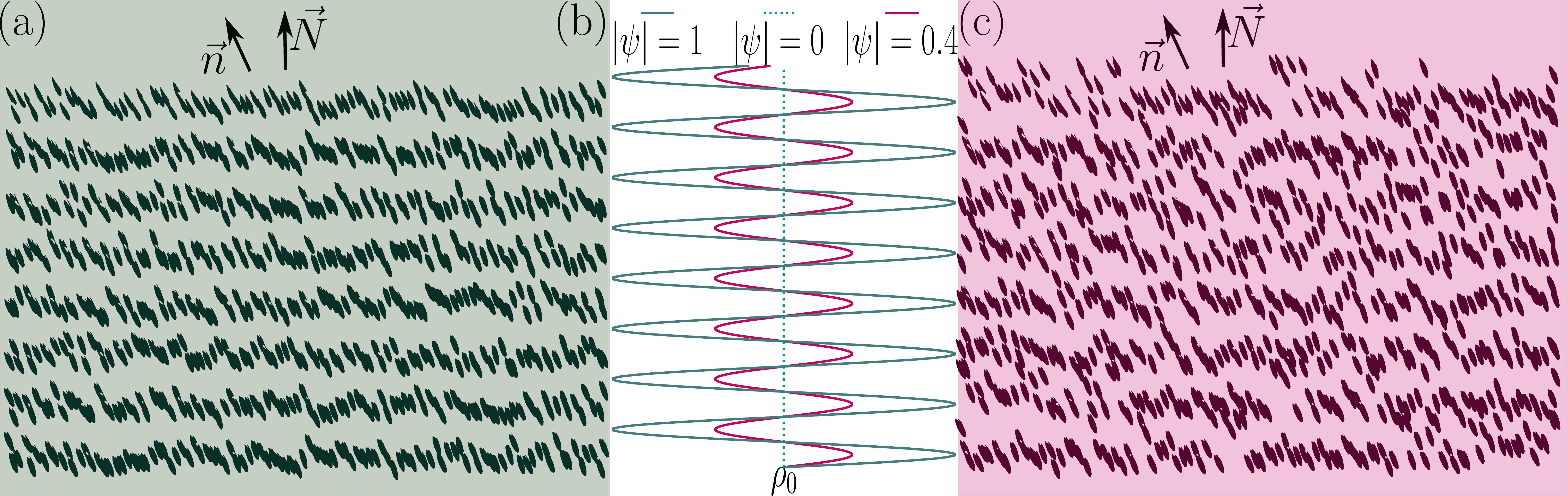}
    \caption{\small
        Randomly generated schematic smectic layers with different values of $\ps$ and associated density variation plots.
        \textbf{(a)} Well ordered layers, $\ps=1$. 
        \textbf{(b)} Plot of $\rho(z)$ for smectic layers aligned in the $z$ direction. The layer normal, $\vec{N}$ points in the $z$ direction and nematic director, $\vec{n}$, aligns with the average direction of the smectogens; 
        \textbf{(c)} Less ordered layers, $\ps=0.4$.
    }
    \label{fig:psi}
\end{figure*}

\begin{figure*}[tb]
    \centering
    \includegraphics[width=0.9\textwidth]{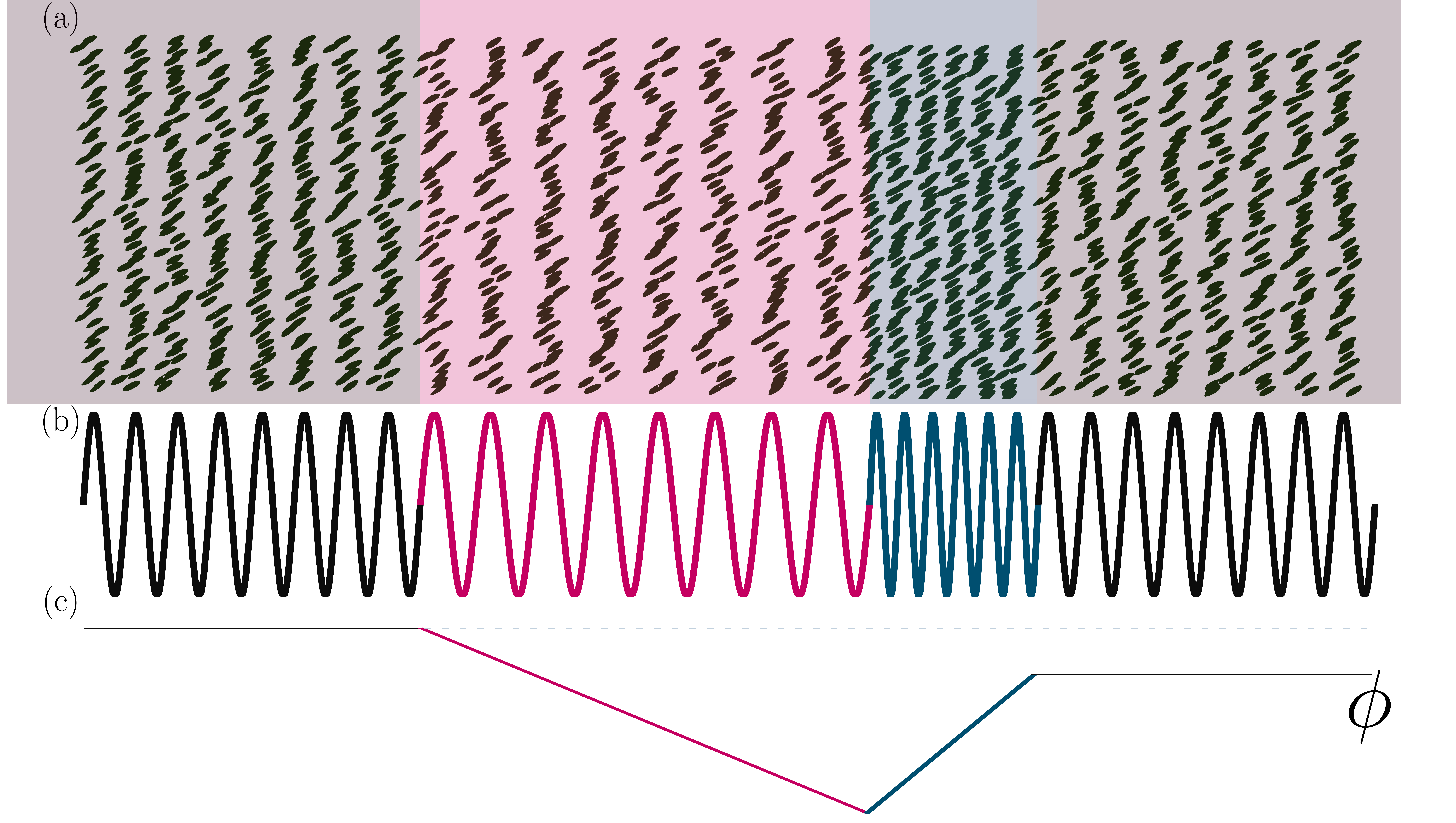}
    \caption{\small
        Illustration of the order parameter: 
        \textbf{(a)} Randomly generated schematic smectic layers showing the effect of linearly changing $\phi$. 
        \textbf{(b)} Plot of $\rho(x)$ for smectic layers aligned in the $x$ direction.
        \textbf{(c)} Plot of corresponding $\phi(x)$. 
    }
    \label{fig:phi}
\end{figure*}

In smectic materials, the smectogen molecules have segregated into well-defined repeating layers. 
Thus, a microscopic description of smectics must begin with the regular oscillation of the average density in the direction of stacked layers. 
To describe smectic layering at time $t$, smectogen density at each point $\vec{r}$ can be approximated by the expansion to lowest order as~\cite{degennes1972}
\begin{align}
	\rho\left(\vec{r},t\right) &= \sum_{m=-\infty}^\infty \psi_m e^{i m \vec{q}\cdot\vec{r}} \approx \psi_{-1} e^{-i \vec{q}\cdot\vec{r}} + \psi_{0} + \psi_{+1} e^{\vec{q}\cdot\vec{r}} + \ldots \nn\\
					&\equiv \rho_0 + 2\RE{ \psi e^{i\vec{q}_0\cdot\vec{r}} }. 
    \label{eq:expansion}
\end{align}
Here, $\rho_0$ is the mean density and the plane wave $\Psi\left(\vec{r},t\right) = \psi e^{i\vec{q}_0\cdot\vec{r}}$ represents the periodic grouping of molecules within layers. 
The assumption of \eq{eq:expansion} is that the density can be approximated sufficiently well by a simple sinusoid $e^{i\vec{q}_0\cdot\vec{r}}$ with any deformations described by $\psi\left(\vec{r},t\right) \in \mathbb{C}$. 
If the lamellae have a well-defined layer normal unit vector $\vec{N}\left(\vec{r},t\right)$ and constant non-perturbed spacing $\thick=2\pi/q_0$, then the wave vector can be written
\begin{align}
	\label{eq:wavevec}
	\vec{q}_0\left(\vec{r},t\right) &= q_0 \vec{N}\left(\vec{r},t\right). 
\end{align}
As proposed by de~Gennes~\cite{degennes1972}, $\psi$ from \eq{eq:expansion} is a complex scalar order parameter, possessing both a modulus and phase
\begin{align}
	\label{eq:orderparam}
	\psi\left(\vec{r},t\right) &= \ps e^{i\phi}. 
\end{align}
The modulus $|\psi\left(\vec{r},t\right)|$  indicates the extent of layering---it is the amplitude of the density wave (\eq{eq:expansion}) that describes accumulation of mesogens into periodic layers (\fig{fig:psi}). 
As $\ps \to 0$, the layers are blurred out and positions become isotropically disordered, representing a phase transition to a isotropic phase. 
By allowing the modulus $\ps$ to vary as a field, regions can locally lose their degree of layering \ie locally melt. 
The phase $\phi\left(\vec{r},t\right)$ measures relative layer displacements---by allowing the phase to vary in space compression/dilation of the density wave can be described. 
Therefore, a global reference for $\phi$ is not meaningful at the macroscopic level but gradients of $\phi$ describe compressional deformations (\fig{fig:phi}). 

According to the phenomenological Ginzburg--Landau theory~\cite{hohenberg2015}, the bulk free energy density $f^\text{bulk}$ can be written as an expansion in terms of the order parameter $\psi$. 
However, the free energy density must be real, $f^\text{bulk} \in \mathbb{R}$, while de~Gennes's order parameter is complex $\psi \in \mathbb{C}$. 
Therefore, the Landau expansion in terms of $\psi$ must have the form
\begin{align}
 \label{eq:bulkTraditional}
 f^\text{bulk}  &= \frac{A}{2} \psi \psi^* + \frac{C}{4} \left( \psi \psi^* \right)^2 + \ldots \nn\\
 			    &= \frac{A}{2} \app + \frac{C}{4}\ps^4  + \ldots
\end{align}
where $\psi^*= \ps e^{-i\phi}$ is the complex conjugate of $\psi$. 
This indicates that the mean field expectation would be a second order phase transition~\cite{chaikin1995}. 
Since $\psi$ is complex, it is analogous to the order parameters for superfluids or superconductors and correspondences between these systems can be physically insightful~\cite{degennes1972,Lubensky1990,navailles2009,Kamien2016,Zappone2020}.
Physically, the phase $\phi$ is absent from \eq{eq:bulkTraditional}  because it represents the relative displacement of layers, which should not enter the bulk free energy. 

The phase $\phi$ is a measure of the layer displacement field $\vec{u}\left(\vec{r},t\right) = u\ \vec{N}\left(\vec{r},t\right)$, where $u=\abs{\vec{u}}$. 
This can seen by writing $\Psi = \psi e^{i\vec{q}_0\cdot\vec{r}} = \ps e^{i\left(\vec{q}_0\cdot\vec{r}+\phi\right)} = \ps e^{i\vec{q}_0\cdot\left(\vec{r}+\vec{u}\right)}$, where 
\begin{align}
	\label{eq:phase}
	\phi\left(\vec{r},t\right) &= \vec{q}_0 \cdot \vec{u} = \left(q_0 \vec{N} \right) \cdot \left( u\vec{N} \right) = q_0 u. 
\end{align}
Thus, the phase measures the relative layer displacement in units of layer spacing $d_0$. 
If $\phi\left(\vec{r},t\right)$ varies, the function in \eq{eq:expansion} is perturbed from being a perfectly periodic sinusoid. 
The value of $\phi$ itself is arbitrary representing a global gauge invariance to where one measures the displacements from --- only variations in $\phi$ are physically meaningful. 
We pause to stress that the planar wave function
\begin{align}
    \Psi\left(\vec{r},t\right) &= \psi e^{i\vec{q}_0\cdot\vec{r}}
    \label{eq:capitalPsi}
\end{align}
explicitly includes the {\em microscopic} spatial information of the density wave down to the scale of layers. 
One the other hand, $\psi = \abs{\psi}e^{i\phi}$ (from \eq{eq:orderparam}) is expected to be a slowly varying parameter that is meaningful on {\em macroscopic} scales, without explicit reference to the  microscopic variation. 

Though elegant, the complex scalar $\psi$-theory has complications and shortcomings~\cite{Pevnyi2014}. 
Fundamentally, $\phi$ is not truly a single-valued function of position and so $\psi$ is not an element of the unit circle $S^1$ but rather the orbifold $S^1 / \mathbb{Z}_2$~\cite{chen2009,alexander2010,alexander2012}. 
Indeed, in this section we have described the layer normal as the vectorial direction of the wave vector; however, an equivalent approach used in the literature is to define the normal as determined directly by phase gradients as $\vec{N} = \del\phi/\abs{\del\phi}$.
% \begin{align}
%     \label{eq:oldN}
%     \vec{N} &= \del\phi/\abs{\del\phi}.
% \end{align} 
As a result, the order parameter does not embed nematic-like rotational symmetry breaking of the layer normal.

%%-------------------------------------------------------------------%%
%%-------------------------------------------------------------------%%
%%-------------------------------------------------------------------%%
%% Nematics and Q-theory %%
%%-------------------------------------------------------------------%%
%%-------------------------------------------------------------------%%
%%-------------------------------------------------------------------%%
\subsection{Q-tensor Landau-de~Gennes theory of nematics}\label{sctn:Q}

Nematic liquid crystals have this same apolar rotational symmetry breaking. 
For these mesophases, the extent of ordering is measured by a real scalar order parameter $S\left(\vec{r},t\right)$ and the local direction is measured by the director field $\vec{n}\left(\vec{r},t\right)$ of unit length. 
In the {\em Ericksen--Leslie} formalism~\cite{ericksen1959, Ericksen:1960, Ericksen:1961, Leslie:1968, leslie1966, borthagaray2020, Zarnescu2021}, the free energy density is constructed from a bulk contribution in terms of $S$ alone and a Frank--Oseen deformation contribution in terms of only gradients of $\vec{n}$. 
However, de~Gennes~\cite{degennes1971,degennes1993} proposed that the order parameter and director can be written together as a tensor order parameter
\begin{align}
 \label{eq:tensQ}
 \tens{Q}\left(\vec{r},t\right) &= S \left( \vec{n} \ \vec{n}-\frac{\tens{\delta}}{d} \right), 
%  \tens{Q}\left(\vec{r},t\right) &= S \left( \vec{n} \ \vec{n}-\tfrac{1}{d}\,\tens{\delta} \right), 
\end{align}
where $\tens{\delta}$ is the identity matrix. 
While one might initially be concerned that a tensorial order parameter could introduce undue mathematical complications, the benefits of $\tens{Q}$-theory are substantial. 
\begin{enumerate}
	\item Firstly, $\tens{Q}$ is even in the director field $\vec{n}$. 
	This immediately ensures that $\vec{n} \to - \vec{n}$ is inconsequential. 
	A tensor formed from the tensor  product $\sim \vec{n} \ \vec{n}$ preserves the desired nematic symmetry, 
	\item The combination of scalar order parameter $S$ and director $\vec{n}$ means that both the bulk free energy and distortion free energy can be written consistently in terms of $\tens{Q}$ as
	\begin{subequations}
	\begin{align}
		f_\text{nem} &= f^\text{bulk}_\text{nem} + f^\text{def} _\text{nem} \label{eq:nem}\\
		f^\text{bulk}_\text{nem} &= \frac{A_\text{nem}}{2} \tens{Q}:\tens{Q} + \frac{B_\text{nem}}{3} \left(\tens{Q}\cdot\tens{Q}\right):\tens{Q} + \frac{C_\text{nem}}{2} \left(\tens{Q}:\tens{Q}\right)^2 + \ldots \nn\\
						 &= \frac{A_\text{nem}}{2} Q_{ij}Q_{ji} + \frac{B_\text{nem}}{3} Q_{ij}Q_{jk}Q_{ki} + \frac{C_\text{nem}}{2} \left(Q_{ij}Q_{ji}\right)^2 + \ldots 
						 \label{eq:nemBulk}
						 \\
		f^\text{def} _\text{nem} &= \frac{L_1}{2} \left( \del \tens{Q} \right)^2 + \frac{L_2}{2} \left(\del\cdot\tens{Q}\right)\cdot\left(\del\cdot\tens{Q}\right) + \frac{L_3}{2} \tens{Q}:\left( \del \tens{Q} \ : \ \del\tens{Q}\right)  \nn\\
							&= \frac{L_1}{2} Q_{ij,k} Q_{ij,k} + \frac{L_2}{2} Q_{kj,k} Q_{ij,i} + \frac{L_3}{2} Q_{ki} Q_{jl,k} Q_{jl,i}. 
							\label{eq:nemDef}
	\end{align}
	\end{subequations}
	If we assume a single elastic coefficient $L_1=L_2=L_3$, the distortion free energy density is simply $f^\text{def} _\text{nem} = L_1 Q_{ij,k}Q_{ij,k}$. 
	These free energies are appealing since all calculations can be done on a single mathematical object $\tens{Q}$. 
	The free energy density was written above first using the vector notation and then Einstein summation convention to be explicit and because we will use both interchangeably, as convenient. 
	Our comma notation is $Q_{ij,k} = \partial_k Q_{ij}$. 
	The elastic constants in \eq{eq:nem}, can be directly related to the Frank coefficients for splay, twist and bend~\cite{schiele1983,shendruk2018}. 
	The free energy density in \eq{eq:nem} amounts to a phenomenological Landau expansion in terms of $\tens{Q}$ and its derivative and so the $\tens{Q}$-theory is refered to as the {\em Landau--de~Gennes} theory~\cite{Andrienko2018,anzivino2020}. 
	\item Setting the constant trace of $\tens{Q}$ to zero is convenient so that the first invariant does not contribute to the free energy. 
	\item The tensor form is {\em numerically} practical. 
	The primary reason for this is because disclination defects exist in many interesting systems. 
	These defects are point singularities in the director field, which are computationally burdensome; however, the inclusion of $S$ allows these singularities to locally melt in a finite defect core, making them more amenable to computational approaches. 
	For this reason, $\tens{Q}$-theory is used extensively in numerical simulations nematics, including colloidal liquid crystals~\cite{beller2015, Boniello2019, hijar2020, villada2021}, living liquid cyrstals~\cite{turiv2020, mandal2021} and active liquid crystals~\cite{Rivas2020, Pearce2020, Pearce2021, zhou2021, thijssen2021}. 
\end{enumerate}

%%-------------------------------------------------------------------%%
%%-------------------------------------------------------------------%%
%%-------------------------------------------------------------------%%
%% Tensor form %%
%%-------------------------------------------------------------------%%
%%-------------------------------------------------------------------%%
%%-------------------------------------------------------------------%%
\section{Smectic Tensorial Order Parameter}\label{sctn:E}

Taking inspiration from the success of $\tens{Q}$-theory, we recently considered a uniaxial, complex, tensorial lamellar order parameter 
\begin{align}
	\label{eq:tensE}
	\boxed{ \tens{E}\left(\vec{r},t\right) = \ps e^{i\phi} \left( \vec{N} \ \vec{N}-\frac{\tens{\delta}}{d} \right) } \ , 
\end{align}
where $\psi\left(\vec{r},t\right) \in \mathbb{C}$ and $\vec{N}\left(\vec{r},t\right) \in \mathbb{R}^d$ for smectic systems; $d$ is the dimensionality of the system~\cite{paget2022}. 
The tensor $\tens{E}$ is analogous to $\tens{Q}$ but the scalar order parameter $\psi=\ps e^{i\phi}$ is complex instead of real, as $S$ is. 
It is a macroscopic object that does not include the microscopic oscilations included in $\Psi$ (see \eq{eq:capitalPsi}). 
It consolidates the complex scalar order parameter $\psi$, the apolar layer normal $\vec{N}$, and resolves the phase ambiguity~\cite{Pevnyi2014}. 
Before developing a phenomenological $\tens{E}$-based Landau theory for smectics (\sctn{sctn:freeen}), we pause to reflect on the origins (\sctn{sctn:arg}), inherent limitations (\sctn{sctn:micro}) and properties (\sctn{sctn:char}) of the complex-tensor order parameter. 

%%-------------------------------------------------------------------%%
%%-------------------------------------------------------------------%%
%%-------------------------------------------------------------------%%
%% Argument %%
%%-------------------------------------------------------------------%%
%%-------------------------------------------------------------------%%
%%-------------------------------------------------------------------%%
\subsection{Argument for Smectic Tensorial Order Parameter}\label{sctn:arg}

First, we expand on the considerations leading us to \eq{eq:tensE}. 
We seek a smectic order tensor $\tens{E}$ that consolidates the complex scalar order parameter $\psi$ and the apolar layer normal $\vec{N}$. 
It should contain the dyadic product of $\vec{N}$ with itself in order for $\vec{N}\to-\vec{N}$ to be inconsequential and so possess up-down symmetry. 
Furthermore, the absence of preferential directions within planar layers means local rotations about $\vec{N}$ should be arbitrary. 
A traceless order parameter ensures linear terms do not contribute to the bulk free energy. 
We are not interested in the microscopic details of individual layers but seek a hydrodynamic scale description --- we prefer a form in which the factor $e^{i\vec{q}_0\cdot\vec{r}}$ does not need to be included, as it represents the microscopic details of individual layers (\sctn{sctn:micro}). 
Instead we want to account for the two pieces of information embedded in de~Gennes' order parameter $\psi = \ps e^{i\phi} \in \mathbb{C}$, where $\ps$ quantifies the extent of layering and $\phi$ represents layer displacements, which is necessary to describe compression/dilation of the layers. 

We assume that the smectic order tensor is diagonalizable, $\diag{\tens{E}}\propto\left[ \lambda_1,\lambda_2,\lambda_3 \right]$ for three complex eigenvalues. 
Given this, $\tens{E}$ can be written in terms of basis vectors $\left(\vec{l}, \vec{m}, \vec{N}\right)$. 
We insist that $\tens{E}$ is traceless so that its first invariant is zero and does not contribute to the bulk free energy. 
We are seeking a form for which at least one of the eigenvalues is proportional to $\psi$. 

Let us next make the spurious assumption that each element of $\tens{E}$ is real. 
In this invalid case, both of the conjugate pair $\psi$ and $\psi^*$ would be eigenvalues. 
The tracelessness of $\tens{E}$ would then require $\diag{\tens{E}}\propto\left[ -\psi^*,-\psi,2\RE{\psi} \right]$. 
This form would be reminiscent of biaxial nematics~\cite{govers1984,Mukherjee2019} in which $\diag{\tens{Q}}=\left[ S-\eta,S+\eta,-2S \right]$ but with the degree of biaxiality $\eta\in\mathbb{R}$ replaced with an imaginary contribution. 
While this analogy might at first appear appealing, we reject this construction because bulk free energy terms, such as $\tens{E}:\tens{E}^*$, would necessarily include contributions from the phase $\phi$, amounting to non-physical excess free energy costs to orientational (\ie nematic) ordering of the layer normals. 

To remove such layer normal orientational contributions, the constraint that $\tens{E}$ is traceless could be relaxed; though this does not fully resolve the issue. 
If one insists that the third eigenvalue is zero, then the nematic alignment of layer normals would not contribute to the bulk free energy expansion; however, $\vec{N}$ would be a nullspace and $\tens{E}$ would be singular. 
Furthermore, gradients of $\tens{E}$ would then include derivatives of the in-plane vectors $\vec{m}$ and $\vec{l}$ and so the distortion free energy would necessarily include biaxial-type contributions~\cite{govers1984}. 
Any such terms must be non-physical because of the isotropic nature of simple smectics within layers. 
Thus, we conclude that $\tens{E}$ is not in general real. 

Following from the in-plane isotropy, $\tens{E}$ should not depend explicitly on either of the arbitrary in-plane unit vectors, $\vec{m}$ or $\vec{l}$. 
Moreover, the physical equivalence of $\vec{m}$ and $\vec{l}$ demands they share the same eigenvalue, just as in a nematic. 
In 3D, tracelessness then demands $\diag{\tens{E}}=\left[ -\psi/3,-\psi/3,2\psi/3 \right]$, which is equivalent to a nematic liquid crystal but with a complex scalar order parameter. 
The tensorial form of $\tens{E}$ for simple lamellar phases then follows from this point by noting that the nematic nature of the layer normal demands that only even dyadic products $\vec{N}\ \vec{N}$ should be present and so we arrive at the proposed form of $\tens{E} = \psi\left( \vec{N}\ \vec{N}-\tens{\delta}/d \right)$ in $d=2,3$ dimensions. 
As a dyadic, the tensorial lamellar order parameter should be symmetric, which means that it cannot be Hermitian. 

For perfectly ordered lamellae in which there is absolutely no layer displacement, one is free to chose $u=0$ such that $\phi=\vec{q}_0\cdot\vec{u}=\left(q_0\vec{N}\right)\cdot\left(u\vec{N}\right)=q_0u=0$, which causes $\psi=\abs{\psi}$ and $\tens{E}$ to be a real, traceless and symmetric tensor. 
However, this is only true for perfectly planarly stacked layers --- all deformations in which the layer normal field $\vec{N}$ varies must be accompanied by some local layer displacement requiring complex $\phi$. 

We now consider an argument that the eigenvectors $\left(\vec{l}, \vec{m}, \vec{N}\right)$ obtained from $\tens{E}$ must be real. 
In particular, the layer normal should be composed of purely real components, $\vec{N}\in\mathbb{R}^d$.
Let $\tens{E}$ be the product of a real symmetric and traceless tensor $\tens{\mathcal{T}}$ and a nonzero complex scalar $\psi$, \ie $\tens{\mathcal{T}} \equiv \vec{N}\ \vec{N}-\tens{\delta}/d$. 
Being real, symmetric and traceless, $\tens{\mathcal{T}}$ will have real eigenvalues, $\tilde{\lambda}_i$, and real orthogonal eigenvectors. 
The complex tensor $\tens{E}$ must share these eigenvectors and have eigenvalues $\lambda_i = \psi\tilde{\lambda}_i$. 
This proof relies on the form $\tens{E}=\psi\tens{\mathcal{T}}$. 
Much like the tracelessness property, this is a condition that must be enforced, in particular in numerical calculations (see \sctn{sctn:lagrange}). 

Though the $\tens{E}$-based theory will be computed in terms of $\tens{E}$ alone, it may be desirable to find $\psi\left(\vec{r},t\right)$ and $\vec{N}\left(\vec{r},t\right)$ after the fact to aid in interpreting results. 
Even though $\psi$ is the eigenvalue of $\tens{E}$ and $\vec{N}$ is the associated eigenvector, eigen-decomposition may not necessarily be the numerically preferred method of determining these quantities. 
For example, it is convenient to find the modulus by contracting $\tens{E}$ with its complex conjugate, 
\begin{align}
    \label{eq:calcModulus}
    \ps^2 &= \frac{\tens{E}:\tens{E}^*}{\varrho} ,
\end{align} 
where $\varrho\equiv\left(d-1\right)/d$. 
To find \eq{eq:calcModulus}, we recognize the layer normal is a unit vector so $\vec{N}\cdot\vec{N}=1$ and $\delta_{ii}=\tr{\tens{\delta}}=d$. 
Similarly, the phase $\phi$ can be found by contracting the complex tensor with itself
\begin{align}
    e^{2i\phi} &= \frac{\tens{E}:\tens{E}}{\varrho\ps^2} \nn\\
    \phi &= (-i/2)\text{arg}\left[ \frac{ \tens{E}:\tens{E} }{\varrho\ps^2}\right] , 
    \label{eq:calcPhase}
\end{align} 
where $\ps$ is known from \eq{eq:calcModulus}.

%%-------------------------------------------------------------------%%
%%-------------------------------------------------------------------%%
%%-------------------------------------------------------------------%%
%% Biaxiality %%
%%-------------------------------------------------------------------%%
%%-------------------------------------------------------------------%%
%%-------------------------------------------------------------------%%
\subsection{Biaxiality}
Here, we have implicitly focused on uniaxial liquid crystals. 
However, in nematics, $\tens{Q}$-tensor theory can be extended to account for biaxiality --- a degree of orientational alignment along a secondary direction. In the biaxial nematic case, we could write
\begin{align}
 \label{eq:tensQbiax}
 \tens{Q}\left(\vec{r},t\right) &= S_1 \left( \vec{n} \ \vec{n}-\frac{\tens{\delta}}{d} \right) + S_2 \left( \vec{m} \ \vec{m}-\frac{\tens{\delta}}{d} \right), 
\end{align}
for scalar order parameters $S_1$ and $S_2$. 
In three dimensions, this represents orthonormal eigenvectors $\vec{n}$, $\vec{m}$, $\vec{l} =\vec{n}\times\vec{m}$\cite{Mucci2017}, corresponding respectively to eigenvalues
\begin{align}
\lambda_1 = \frac{2S_1-S_2}{3} \qquad ; \qquad
\lambda_2 &= \frac{2S_2-S_1}{3} \qquad ; \qquad
\lambda_3 = -\frac{S_1+S_2}{3}. 
\end{align}
In this biaxial nematic case, the number of degrees of freedom is increased from the uniaxial case --- increasing the degrees of freedom from three to five in three dimensions.

It is interesting to ask if a biaxiality in smectic ordering could be possible, which might describe secondary layering in an orthonormal direction. 
The uniaxial smectic order parameter tensor defined in \eq{eq:tensE} is complex valued symmetric and traceless. 
Drawing analogy to the biaxial $\tens{Q}$-tensor in \eq{eq:tensQbiax}, we could also consider 
\begin{align}
 \label{eq:tensEbiax}
 \tens{E}\left(\vec{r},t\right) &= \psi_1 \left( \vec{N} \ \vec{N}-\frac{\tens{\delta}}{d} \right) + \psi_2 \left( \vec{m} \ \vec{m}-\frac{\tens{\delta}}{d} \right),
\end{align}
for complex scalar order parameters $\psi_1$ and $\psi_2$ and real orthonormal eigenvectors $\vec{N}$, $\vec{m}$, $\vec{l} =\vec{N}\times\vec{m}$. 
The biaxial smectic order parameter tensor gains two additional degrees of freedom compared to the biaxial $\tens{Q}$ simply because the eigenvalues are complex, representing the degree of ordering and layer displacement in the secondary direction. 
Physically, the biaxial tensor would begin to describe a degree of layering in a second direction --- not necessarily at the same spacing or extent as the primary direction. 
The main practical difference between these two expressions for $\tens{E}$ is that the biaxial case will not necessarily have a uniform complex phase across different components of the tensor. 
This would not be equivalent to a biaxial sm-A\textsubscript{b}~\cite{Meyer2021} but might, for example, reflect secondary layering in colloidal banana-shaped dispersions~\cite{rico2020}.

%%-------------------------------------------------------------------%%
%%-------------------------------------------------------------------%%
%%-------------------------------------------------------------------%%
%% Micro %%
%%-------------------------------------------------------------------%%
%%-------------------------------------------------------------------%%
%%-------------------------------------------------------------------%%
\subsection{Approximate Microscopic Form}\label{sctn:micro}

Our tensorial order parameter is concerned with the complex scalar order parameter $\psi=\abs{\psi}e^{i\phi}$; rather than the full plane wave $\Psi=\psi e^{i\vec{q}_0\cdot\vec{r}}$. 
The order parameter $\tens{E}\left(\vec{r},t\right)$ is assumed to be a {\em hydrodynamic variable} that varies on length and time scales that are large compared to microscopic scales. 
This reflects our intent to construct a theoretical description of the macroscopic state of simple smectics, through the density modulation $2\ps$ and the equilibrium wave number $q_0=2\pi/d_0$ for layer spacing $d_0$. 
The wave vector combines the wave number and layer normal $\vec{q}_0=q_0\vec{N}$ --- the direction can vary on length and time scales, such that microscopically the plane wave remains a satisfactory approximation of the smectogen density. 
We regard the local distribution of smectogens within layers and description of individual layers themselves to belong to the microscale.  

Since this construction assumes a local microscopic planar wave for the density of smectogens, the microscopic structure can be directly reproduced from $\tens{E}$ within an arbitrary global phase shift. 
Just like the traditional smectic scalar order parameter $\psi=\abs{\psi}e^{i\phi}$ can be interpreted as the complex amplitude of the density oscillations $\Psi=\psi e^{i\vec{q}_0\cdot\vec{r}}$~\cite{degennes1972}, $\tens{E}$ is the complex tensorial amplitude of 
\begin{align}
 %\tens{\mathcal{E}} &= \Psi\left( \vec{N}\ \vec{N}-\frac{\tens{\delta}}{d} \right)
  \tens{\mathcal{E}} &= \Psi\left( \vec{N}\ \vec{N}-\tfrac{1}{d}\,\tens{\delta} \right)
     = \tens{E} e^{iq_0\vec{N}\cdot\vec{r}}. 
     \label{eq:microE}
\end{align}
In this work, we employ $\tens{\mathcal{E}}$ sporadically, when convenient, in particular when making contact with descriptions that utilize covariant derivatives or determining microscopic equilibrium values (\sctn{sctn:eq}). 
The microscopic form $\tens{\mathcal{E}}$ should be understood to only contain the same information as the macroscopic form but with the assumed microscopic density wave from \eq{eq:expansion} locally overlayed. 

%%-------------------------------------------------------------------%%
%%-------------------------------------------------------------------%%
%%-------------------------------------------------------------------%%
%% Character %%
%%-------------------------------------------------------------------%%
%%-------------------------------------------------------------------%%
%%-------------------------------------------------------------------%%
\subsection{Characteristics of Smectic Tensorial Order Parameter}\label{sctn:char}
We describe the tensorial smectic order parameter $\tens{E}$ as complex, traceless, normal, globally gauge invariant and symmetric but non-Hermitian. 
\begin{description}
    \item[Complex] As a complex tensor in $d$ dimensions, $\tens{E}$ contains $2d^2$ elements. As explained in \sctn{sctn:arg}, one could take the layer-displacement to be $0$, in which case $\tens{E}$ would become real (setting $u=0$). However, any deformation introduced to the material with respect to the layer normal must be accompanied by a phase gradient, hence introducing a non-zero imaginary part to components of $\tens{E}$.
    \item[Traceless] Analogous to the nematic $\tens{Q}$-tensor, $\tens{E}$ is chosen to be traceless
    \begin{subequations}
    \begin{align}
        \label{eq:traceless}
        \tr{\tens{E}} &= E_{ii} = 0. 
    \end{align}
    A constant trace is equivalent to the constraint that the rod-shaped smectogen particles are of fixed length. The constant trace is fixed to zero to ensure linear terms do not contribute to the bulk free energy. 
    \item[Normal Operator] $\tens{E}$ is a normal tensor, meaning it commutes with its Hermitian adjoint $\tens{E}^*$, \ie 
    \begin{align}
        \label{eq:normal}
        \commutator{ \tens{E} }{ \tens{E}^* } &= \tens{0} , 
    \end{align}
    where $[A,B]=AB-BA$ is a commutator. 
    % In terms of \eq{eq:tensE}, this means the $d^2$ components of $\tens{E}$ are required to have the same complex phase. 
    % \end{subequations}
    %
    \item[Uniaxial]
    In our study, we fix $\tens{E}$ to the uniaxial case as defined in \eq{eq:tensE}. Practically, this is equivalent to constraining the $d^2$ components of $\tens{E}$ to have the same complex phase. We can express this as 
    \begin{align}
        \label{eq:uni_const}
        \det{ \commutator{ \tens{E} }{ \tens{E}^* } } &= \tens{0} , 
    \end{align}
    This comes from $\det{[\tens{E},\tens{E}^*]}$ being proportional to $\text{sin}^2(\Theta)$ for $\Theta=\text{arg}(E_{ii})-\text{arg}(E_{ij})$ for $i\ne j$. In two dimensions, this is equivalent to \eq{eq:normal}.
    \end{subequations}
    \item[Globally Gauge Invariant] As can be seen in \fig{fig:phi}, the specific value of $\phi$ is arbitrary and only gradients, representing compression or dilation, are physically meaningful. 
    Hence, we can perform the gauge transformation $\tens{E} \to e^{i\alpha}\tens{E}$, where $\alpha$ is a global constant, which only reflects a constant global shift for the system; rather than a physically meaningful transformation. 
    Any proposed free energy (\sctn{sctn:freeen}) and thermodynamic observables are invariant under this gauge transformation. 
    \item[Symmetric and non-Hermitian] Writing $\tens{E} = \psi\left( \vec{N}\ \vec{N}-\tens{\delta}/d \right) \equiv \psi \tens{\mathcal{T}}$ makes it clear that $\tens{\mathcal{T}}=\vec{N}\ \vec{N}-\tens{\delta}/d$ is symmetric. 
    However $\tens{E}$ is not Hermitian, as $E_{ij}=E_{ji}$ rather than $E_{ij}=E_{ji}^*$.
\end{description}
Given these characteristics and constraints, there are $d+1$ degrees of freedom embedded in the complex order parameter tensor $\tens{E}$. 
This is one more than are embedded in the uniaxial nematic $\tens{Q}$ tensor and they represent the physical extent of layering, layer displacement and direction. 

%%-------------------------------------------------------------------%%
%%-------------------------------------------------------------------%%
%%-------------------------------------------------------------------%%
%% Free energy %%
%%-------------------------------------------------------------------%%
%%-------------------------------------------------------------------%%
%%-------------------------------------------------------------------%%
\section{Landau Free energy}\label{sctn:freeen}

Having proposed a complex-tensor order parameter for lamellar materials, we now construct a model for the free energy in terms of $\tens{E}$. 
Though smectics and other lamellae have been modelled from many perspectives~\cite{emelyanenko2015,alageshan2017}, we consider a phenomenological Landau free energy expansion. 
The total free energy
\begin{align}
 \label{eq:totalFreeEn}
 F(t) &= \int d^d\vec{r} \ f\left(\tens{E},\del \ \tens{E}\right), 
\end{align}
is an integral over space of a free energy density $f\left(\vec{r},t\right)$. 
Once identified, Landau theory expands the free energy density in terms of the order parameter and its derivatives. 
The expansion must be stable against unlimited growth of the order parameter. 
This corresponds to the Landau--de~Gennes theory for nematics (\sctn{sctn:Q}), except that the order parameter tensor $\tens{E}$ is complex. 
In the case where $\phi$ is uniform in space, $\tens{E}$ is equivalent to $\tens{Q}$ times a complex constant, though the free energy should include different terms.
The form of the free energy must make replacing $\tens{E}\to\tens{E}^*$ immaterial, reflecting the double-valued nature of $\psi=\abs{\psi}e^{i\phi}$~\cite{Pevnyi2014}. 
This global transformation would be physically equivalent to switching all regions of compression to dilation and vice versa, though the boundary conditions would distinguish the two as distinct systems. Under large deformation, this equates to an equal cost to either inserting or removing a layer. 
Thus, we expect all terms in the free energy to involve pairings of $\tens{E}$ and its complex conjugate $\tens{E}^*$. 
Furthermore, with both a translational and rotational symmetry breaking, a tensor-based framework for smectic liquid crystals must account for {\em at least} two distinct elastic constants~\cite{chaikin1995}. 

As in Landau--de~Gennes theory, we group the free energy density expansion into bulk and deformation contributions. 
We consider five contributions to the total free energy of a simple lamellar fluid
\begin{align}
 \label{eq:freeen}
 F &= \int d^d\vec{r}  \left[ \ f^\text{bulk}\left(\tens{E}\right) + f^\text{comp}\left(\del \ \tens{E}\right) + f^\text{curv}\left(\del^2\tens{E}\right) + f^\text{ext}\left(\tens{E}\right) \right] \nn\\ 
    &\qquad +  \int dS \ \mathcal{F}^\text{anch}\left(\tens{E}\right), 
\end{align}
which is the sum of the volume integral over the bulk ($f^\text{bulk}$), two deformation (layer compression/dilation $f^\text{comp}$ and layer bending $f^\text{curv}$) and external field ($f^\text{ext}$) free energy densities and the surface contribution, which represents the surface anchoring free energy per unit area ($\mathcal{F}^\text{anch}$). 
This is the free energy of only the layers themselves --- it does not include any free energy contribution due to the alignment of smectogen molecules within the layers, as would be the case in Sm-A or Sm-C thermotropic smectic liquid crystals. 
Accounting for this alignment would require coupling \eq{eq:freeen} to \eq{eq:nem}, similar to the coupling of $\tens{Q}$ to the real-valued variationof density in Ref.~\cite{xia2021}. 
Thus, the current, non-coupled theory is directly relevant to lyotropics. 

%%-------------------------------------------------------------------%%
%%-------------------------------------------------------------------%%
%%-------------------------------------------------------------------%%
%% Sm Bulk Free Energies %%
%%-------------------------------------------------------------------%%
%%-------------------------------------------------------------------%%
%%-------------------------------------------------------------------%%
\subsection{Smectic Bulk Terms} 
Since $\tens{E}$ is traceless, the bulk smectic free energy density can be written as an even expansion
\begin{align}
	\label{eq:bulkSm}
	\boxed{ f^\text{bulk} = \frac{A}{2}E_{ij}E_{ji}^* + \frac{C}{4}\left(E_{ij}E_{ij}^*\right)^2+\ldots } \ ,
\end{align}
where $C>0$ and $A<0$ produces lamellar order but $A>0$ does not.  
This is the same form as stated in Ref.~\cite{paget2022}. 
The bulk free energy does not depend on phase or layer normal direction, only the complex scalar order parameter. 
This can be seen by explicitly substituting \eq{eq:tensE} into \eq{eq:bulkSm} and recalling 
% \begin{align}
%  \label{eq:E:E}
%  E_{ij}E_{ji}^* 
%                 % &= \ps e^{i\phi} \left( N_iN_j-\frac{\delta_{ij}}{d} \right) \ps e^{-i\phi} \left( N_jN_i-\frac{\delta_{ji}}{d} \right) \nn\\
%                 %= \app \left( N_iN_jN_jN_i - N_iN_j\frac{\delta_{ji}}{d} -\frac{\delta_{ij}}{d}N_jN_i + \frac{\delta_{ij}}{d}\frac{\delta_{ji}}{d} \right)\nn\\
%                 % = \app \left( 1 - \frac{2}{d} + \frac{d}{d^2} \right) 
%                  &= \varrho \app
% \end{align}
$E_{ij}E_{ji}^* = \varrho \app$ from \eq{eq:calcModulus}. 
Then the bulk free energy is simply 
\begin{align}
	\label{eq:bulkTrad}
	f^\text{bulk} &= \frac{A}{2}\varrho \app + \frac{C}{4}\varrho^2 \ps^4 . 
\end{align}
This form matches the form of \eq{eq:bulkTraditional} if $\varrho$ is absorbed into the coefficients and demonstrates the consistency between the complex tensor theory approach and traditional bulk free energies for smectics~\cite{degennes1972,renn1988}. 
In the mean-field limit, \eq{eq:bulkSm} predicts a second order phase transition from isotropic to lamellae. 
The bulk contribution in \eq{eq:bulkTrad} forms the foundation of free energy expansions in terms of the traditional scalar order parameter for the transitions to a wide number of smectic phases~\cite{Mukherjee2021b, mukherjee2001, mukherjee2005, mukherjee2007, das2008, das2009, Mukherjee2019, kats2019, Mukherjee2021a}

%%-------------------------------------------------------------------%%
%%-------------------------------------------------------------------%%
%%-------------------------------------------------------------------%%
%% Sm Compression %%
%%-------------------------------------------------------------------%%
%%-------------------------------------------------------------------%%
%%-------------------------------------------------------------------%%
\subsection{Smectic Layer Compression Terms}

The bulk free energy from \eq{eq:bulkSm} is rarely the only contribution in smectic materials, even after extended relaxation towards equilibrium. 
This is because lamellar phases very rarely reach the equilibrium configuration of perfectly aligned planar layers~\cite{Hur2015,kim2019,hur2018,rottler2020}. 
Rather, they often form defect-populated quasi-phases. 
These include focal conics~\cite{Liarte2015, kim2009a, kim2009b, kim2010a, kim2010b, gim2017, suh2019, preusse2020} and glassy configurations of defect-pinned domains~\cite{hou1997,boyer2002}, which are the smectic equivalent of Shubnikov phases in type-II superconductors~\cite{degennes1972}. 

We first consider compression and dilation of the layers, which we refer to jointly as {\em compression}. 
The compression free energies involve derivatives of the tensor order parameter and must be guaranteed to be real. 
The simplest such term is $E_{ij,k} E_{ij,k}^*$, where $k$ denotes the Cartesian direction over which the gradient is taken~\cite{paget2022}. 
Additional real terms could be constructed through combinations of similar terms and their complex conjugates, which would allow different deformation modes to possess differing elastic moduli at the cost of complicating the theory. 
Most appreciably, deformations parallel and perpendicular to the layer normal should contribute separately to the free energy. 
The compression/dilation free energy is thus taken to be
\begin{align}
    \label{eq:comp}
	\boxed{ f^\text{comp} = b_1^\parallel \Pi_{k\ell} E_{ij,k} E_{ij,\ell}^* + b_1^\perp T_{k\ell} E_{ij,k} E_{ij,\ell}^*  } \ ,
\end{align}
where  $b_1^\parallel$ is the layer compression elastic modulus, $b_1^\perp$ is a stretching elastic modulus, and $\tens{\Pi}$ and $\tens{T}$ are projection operators. 
Projection operators have similarly been applied to scalar models of smectics~\cite{chen1976,renn1988,renn1992,lukyanchuk1998,kundagrami2003,park2006, joo2007,ogawa2007} and, even when projection operators are not explicitly employed, different coefficients can be used for gradients in different directions~\cite{chen2013}. 

%%-------------------------------------------------------------------%%
%% Projection %%
%%-------------------------------------------------------------------%%
\subsubsection{Projection Operators}\label{sctn:projections}
The deformations have been broken into separate contributions {\em parallel} and {\em perpendicular} to the layer normal. 
This is done through two projection operators
\begin{subequations}
  \label{eq:proj}
  \begin{align}
   \tens{\Pi} &= \vec{N} \ \vec{N} \\
   \tens{T} &= \tens{\delta} - \tens{\Pi} . 
   \label{eq:projPerp}
  \end{align}
\end{subequations}
The projection operators parallel $\tens{\Pi}$ and perpendicular $\tens{T}$ to the layer normal both involve only the outer product of the layer normal with itself, and so they maintain the nematic-type symmetry of invariance under $\vec{N}\to-\vec{N}$. 
However, in the present complex tensorial framework, $\tens{E}\left(\vec{r}\right)$ is computed explicitly, while $\vec{N}$ is found \textit{ex post facto}. 
Therefore, we wish to define projection operations in terms of $\tens{E}$. 
To this end, we note that 
\begin{align}
 \tens{E}\cdot\tens{E}^* 
                        %     &= E_{ij}E_{jk}^*
                        %   = \ps e^{i\phi} \left( N_iN_j-\frac{\delta_{ij}}{d} \right) \ps e^{-i\phi} \left( N_jN_k-\frac{\delta_{jk}}{d} \right) \nn\\
                         % &= \app \left( N_iN_jN_jN_k - N_iN_j\frac{\delta_{jk}}{d} -\frac{\delta_{ij}}{d}N_jN_k + \frac{\delta_{ij}}{d}\frac{\delta_{jk}}{d} \right)\nn\\
                        %  &= \app \left( N_iN_k - \frac{2N_iN_k}{d} + \frac{\delta_{ik}}{d^2} \right)
                         &= \frac{\app}{d} \left( (d-2) \vec{N} \ 
                          \vec{N} + \frac{ \tens{\delta}}{d} \right)\nn\\
 N_iN_k &= \frac{1}{d-2}\left[ \frac{d}{\app}E_{ij}E_{jk}^* - \frac{\delta_{ik}}{d}\right]. 
\end{align}
Therefore through \eq{eq:calcModulus} for $\app$, we define the $\tens{E}$-based parallel projection operator to be
\begin{align}
 \tens{\Pi} &\equiv \frac{d-1}{d-2}\left[ \frac{\tens{E}\cdot\tens{E}^*}{ \tens{E}:\tens{E}^* } - \frac{\tens{\delta}}{d^2\varrho}\right] . 
 \label{eq:projParallel}
\end{align}
As is clear from the denominator of the coefficient in \eq{eq:projParallel}, another approach is required for $d=2$ dimensional systems. 
We must consider the form of the tensor prior to being made traceless $\tens{E}+\psi\tens{\delta}/d = \psi \vec{N} \ \vec{N}$, which is directly proportional to $\tens{\Pi}$ but does require that the complex $\psi$ be found, rather than the simpler and real $\ps$.
However, this can be done via \eq{eq:calcPhase}, allowing us to state 
\begin{align}
 \psi &= \left(\app e^{2i\phi}\right)^{1/2} = \left(\frac{\tens{E}:\tens{E}}{\varrho}\right)^{1/2}. 
\end{align}
Hence, the parallel projection operator in terms of the complex order parameter tensor is
\begin{align}
\label{eq:PI_def}
 \tens{\Pi} &\equiv \frac{\tens{E}}{\psi} + \frac{\tens{\delta}}{d}
            = \left(\frac{\varrho}{\tens{E}:\tens{E}}\right)^{1/2}\tens{E} + \frac{\tens{\delta}}{d}. 
\end{align}
We expect the presence of $\psi$ in \eq{eq:PI_def} to make \eq{eq:projParallel} slightly preferable when $d>2$. 
The perpendicular projector $\tens{T}$ follows directly from \eq{eq:projPerp} and, hereafter, we use the $\tens{E}$-based projection operators. 
Approaches using projection operators for surface anchoring in nematic systems commonly simplify these forms by neglecting variations in the scalar order parameter and replacing it with its constant, equilibrium value~\cite{ravnik2009}. 

%%-------------------------------------------------------------------%%
%%-------------------------------------------------------------------%%
%%-------------------------------------------------------------------%%
%% Sm Curvature Distortion %%
%%-------------------------------------------------------------------%%
%%-------------------------------------------------------------------%%
%%-------------------------------------------------------------------%%
\subsection{Smectic Layer Curvature Terms}
In addition to compression deformations, the shape of the layers can be deformed from perfectly flat planar layers. 
% The amount to bending deformations of the layers. 
Accounting for the elastic cost of the local induced curvatures in the free energy density is common in theories for bending bilayers~\cite{Helfrich1994}. 
To describe the curvature requires second order derivatives of the layer displacement. 
Similar to the compression/dilation free energy, the free energy cost of bending distortions of the layer can be decomposed into contributions parallel and perpendicular to the layer normal. 
Again, we keep only the simplest possible terms: purely parallel, purely perpendicular and a mixed contribution, which we write as
\begin{equation}
	\label{eq:curv}
    \boxed{
    	f^\text{curv} = b_2^\parallel \Pi_{k\ell}E_{ij,\ell k} \Pi_{mn}E_{ij,nm}^* + b_2^\perp T_{k\ell}E_{ij,\ell k} T_{mn}E_{ij,nm}^* + b_2^{\parallel\perp} \left(\Pi_{k\ell}E_{ij,\ell k} T_{mn}E_{ij,nm}^* +T_{k\ell}E_{ij,\ell k} \Pi_{mn}E_{ij,nm}^* \right) \ , 
    }
\end{equation}
% \begin{align}
% 	\label{eq:curv}
% 	\boxed{f^\text{curv} = b_2^\parallel \Pi_{k\ell}E_{ij,\ell k} \Pi_{mn}E_{ij,nm}^* + b_2^\perp T_{k\ell}E_{ij,\ell k} T_{mn}E_{ij,nm}^*  + b_2^{\parallel\perp} \left(\Pi_{k\ell}E_{ij,\ell k} T_{mn}E_{ij,nm}^* +T_{k\ell}E_{ij,\ell k} \Pi_{mn}E_{ij,nm}^* \right)} \ , 
% \end{align}
where $b_2^\parallel$, $b_2^\perp$ and $b_2^{\parallel\perp}$ are elastic constants, and $E_{ij,\ell k} \equiv \partial_\ell \partial_k E_{ij}$. 
The projection operators act on derivatives $f^\text{comp}$ and the symmetry between $\tens{E}$ and $\tens{E}^*$ enforces the single elastic constant on the final two terms of \eq{eq:curv}. 

We refer to this term as the ``curvature'' free energy to make contact with Canham–-Helfrich elasticity theory for membranes, which describes the bending of bilayers in terms of the mean and Gaussian curvatures~\cite{Helfrich1994}. 
However, we avoid the term ``bending'' free energy to avoid any potential ambiguity with nematic bending deformations in the Frank--Oseen formalism of \eq{eq:nemDef}. 
While \eq{eq:curv} shares the form of generalized smectic scalar models~\cite{ogawa2007}, below in \sctn{sctn:reducing} we will demonstrate that the curvature contribution simplifies significantly under appropriate approximations.

%%-------------------------------------------------------------------%%
%%-------------------------------------------------------------------%%
%%-------------------------------------------------------------------%%
%% External field %%
%%-------------------------------------------------------------------%%
%%-------------------------------------------------------------------%%
%%-------------------------------------------------------------------%%
\subsection{Smectic Coupling to an External Field}

While perhaps nonphysical, external field-induced lamellar-ordering can be included through a contribution that is second-order in both a vectorial external field $\vec{H}$ and the smectic order parameter 
\begin{align}
	\label{eq:ext}
	f^\text{ext} &= \chi \vec{H}\cdot\left(\tens{E}\cdot\tens{E}^*\right)\cdot\vec{H}
			= \chi H_{i} E_{ij} E_{jk}^* H_{k},  
\end{align}
where $\chi$ is a susceptibility coefficient. 

%%-------------------------------------------------------------------%%
%%-------------------------------------------------------------------%%
%%-------------------------------------------------------------------%%
%% Surface anchoring %%
%%-------------------------------------------------------------------%%
%%-------------------------------------------------------------------%%
%%-------------------------------------------------------------------%%
\subsection{Smectic Surface Anchoring Terms}
We have now considered each of the contributions to the free energy within the volume integral in \eq{eq:freeen}. 
If the smectic is contained by walls or otherwise in contact with surfaces, a surface contribution to the free energy is required to model the anchoring of the smectic to the walls. 
In our previous work~\cite{paget2022}, we assumed an infinitely strong anchoring condition. 
Here, we relax that. 
We present three forms of the free energy per unit area $\mathcal{F}^\text{anch}$, which respectively describe the cases (i) when the layer normal is anchored to the {\em surface normal} direction ({\em Rapini--Papoular} anchoring), (ii) when the layer normal is anchored to a {\em specific} easy direction that does not correspond to the surface normal and (iii) when the layer normal is forced to lie in a plane but is free to take any orientation within the surface ({\em Fournier} degenerate anchoring). 
Conveniently, both of the equivalent cases for nematics are already even in $\tens{Q}$~\cite{ravnik2009}. 
We then consolidate these into two generic forms. 

Since the nematic Rapini--Papoular surface anchoring free energy per unit area is already even~\cite{ravnik2009}, it is straightforward to generalize to a lamellar version
\begin{align}
 \label{eq:anchRP}
 \mathcal{F}^\text{anch}_\text{RP}  &= \frac{W_0}{2}\left( \tens{E}-\tens{E}^\text{w}\right) : \left( \tens{E}-\tens{E}^\text{w}\right)^* 
 							= \frac{W_0}{2}\left( E_{ij}-E_{ij}^\text{w}\right)  \left( E_{ji}-E_{ji}^\text{w}\right)^* , 
\end{align}
for anchoring to some wall-specified $\tens{E}^\text{w}= \psi^\text{w} e^{i\phi^\text{w}} \left( \vec{\nu}^\text{w} \vec{\nu}^\text{w} - \tens{\delta} / d \right)$. 
This represents a quadratic free energy penalty to deviations from the preferred complex tensorial value. 
The wall-specific $\tens{E}^\text{w}$ not only sets an easy direction for alignment $\vec{\nu}^\text{w}$, but also a preferred complex amplitude $\psi^\text{w}$ --- both degree of ordering $\ps^\text{w}$ and also a surface-preferred phase $\phi^\text{w}$. 
Such a boundary condition might be chosen if smectic layers stack in-plane with the wall surface for instance. 
In this case, the layer normal prefers to be parallel to the surface normal $\vec{\nu}^\text{w}$, the presence of the wall encourages a well-defined layer value of $\ps^\text{w}$, and steric interactions resist layer displacement, setting a preferred phase $\phi^\text{w}$, equivalent to demanding a density trough at the wall. 

However, Rapini--Papoular anchoring may not be appropriate for all boundaries. 
Consider, for instance, smectic layers coming flush to the surface, \ie with the layer normal $\vec{N}$ aligned with an in-plane easy direction $\vec{\nu}^\text{w}$. 
In this case, $\ps^\text{w}$ may or may not have a preferred value at the surface but it is unlikely that the phase $\phi$ is anchored. 
Thus, we separately anchor the layer normal $\vec{N}$ and the degree of layering $\ps$ to the wall values through
\begin{align}
    \label{eq:anchEasy}
    \mathcal{F}^\text{anch}_\text{easy} &= \frac{W_1}{4} \left( \tens{\Pi}-\tens{\Pi}^\text{w} \cdot \tens{\Pi} \cdot \tens{\Pi}^\text{w} \right) : \left( \tens{\Pi}-\tens{\Pi}^\text{w} \cdot \tens{\Pi} \cdot \tens{\Pi}^\text{w} \right) + \frac{W_2}{4} \left( \ps^2 - {\ps^\text{w}}^2 \right)^2,
\end{align}
with $\tens{\Pi}$ as in \eq{eq:PI_def} and $\tens{\Pi}^\text{w}=\vec{\nu}^\text{w} \vec{\nu}^\text{w}$ is the projection operator for the easy alignment direction $\vec{\nu}^\text{w}$. 
The first term favours the layer normal aligning with the easy direction and has anchoring strength $W_1$, while the second term (which can be written as in \eq{eq:calcModulus}) pushes the smectic order towards the surface-preferred degree of order $\ps^\text{w}$ and the phase is free. 

While the Rapini--Papoular form (\eq{eq:anchRP}) works well to set the layer normal uniformally along a given easy axis, surfaces with planar anchoring that lack a preferred in-plane direction require the degenerate form~\cite{Fournier2005}. 
The degenerate form can again follow nematic theory, which is given by the Fournier surface anchoring free energy per unit area
\begin{align}
    \label{eq:anchDeg}
     \mathcal{F}^\text{anch}_\text{deg}  &= \frac{W_1}{4} \left( \tens{\Pi}-\tens{T}^\text{w} \cdot \tens{\Pi} \cdot \tens{T}^\text{w} \right) : \left( \tens{\Pi}-\tens{T}^\text{w} \cdot \tens{\Pi} \cdot \tens{T}^\text{w} \right) + \frac{W_2}{4} \left( \ps^2 - {\ps^\text{w}}^2 \right)^2 , 
\end{align}
where $\tens{T}^\text{w}=\tens{\delta} - \vec{\nu}^\text{w}\vec{\nu}^\text{w}$ is the surface projection for a surface with normal $\vec{\nu}^\text{w}$. 
As in \eq{eq:anchEasy}, the first term favours the layer normal lying in the plane of the surface, while the second term penalizes degrees of order that differ from $\ps^\text{w}$. 
Unlike the typically employed nematic form (see Refs.~\cite{ravnik2009,gim2017}), the first term of $\mathcal{F}^\text{anch}_\text{deg}$ in \eq{eq:anchDeg} is based on $\tens{\Pi}$ (\eq{eq:projPerp}) instead of the traceless version that includes a factor of the scalar order parameter~\cite{ravnik2009} and so acts solely on the direction of the layer normal and does not anchor the degree of ordering to any particular value. 

The specific easy-direction (\eq{eq:anchEasy}) and degenerate (\eq{eq:anchDeg}) anchoring free energies can be consolidated and generalized if we recognize that $\tens{\Pi}^\text{w}=\vec{\nu}^\text{w} \vec{\nu}^\text{w}$ and $\tens{T}^\text{w}=\tens{\delta} - \vec{\nu}^\text{w}\vec{\nu}^\text{w}$ are just two possible examples of a wall projection operator $\tens{P}^\text{w}$. 
Furthermore, the Fournier form (\eq{eq:anchDeg}) sets $\vec{N}$ and $\ps$ with different anchoring strengths, whereas the Rapini--Papoular form (\eq{eq:anchRP}) imposes equal anchoring on $\vec{N}$, $\ps$ and $\phi$. 
Thus, we generalize all these cases to 
\begin{align}
    \label{eq:anchGen}
     \mathcal{F}^\text{anch}  &= \frac{W_1}{4} \left( \tens{\Pi}-\tens{P}^\text{w} \cdot \tens{\Pi} \cdot \tens{P}^\text{w} \right) : \left( \tens{\Pi}-\tens{P}^\text{w} \cdot \tens{\Pi} \cdot \tens{P}^\text{w} \right) \nn\\
        &\qquad + \frac{W_2}{4} \left( \ps^2 - {\ps^\text{w}}^2 \right)^2  + \frac{W_3}{4} \left( e^{i\phi}e^{i\phi^\text{w}} + \left[e^{i\phi}e^{i\phi^\text{w}}\right]^* \right)^2 , 
\end{align}
where the last term is $\sim\cos^2\left(\phi-\phi^\text{w}\right)$ and $e^{i\phi}$ can be conveniently found via \eq{eq:calcPhase}.

Having constructed a Landau free energy for the lamellar system in terms of the smectic complex-tensor $\tens{E}$, we now consider the physical ramifications of this model. 
From this point on, we neglect external and surface contributions to the free energy and consider the total free energy density to be the sum of \eq{eq:bulkSm}, \eqref{eq:comp} and \eqref{eq:curv}
\begin{align}
\label{eq:total}
	f &= f^\text{bulk} + f^\text{comp} + f^\text{curv}  \nn\\
	&= \frac{A}{2}E_{ij}E_{ji}^* + \frac{C}{4}\left(E_{ij}E_{ij}^*\right)^2 
	+ \left[ b_1^\parallel \Pi_{k\ell}  + b_1^\perp T_{k\ell} \right] E_{ij,k} E_{ij,\ell}^* \nn\\
	&\quad + \left[ b_2^\parallel \Pi_{k\ell} \Pi_{mn} + b_2^\perp T_{k\ell}T_{mn}  + b_2^{\parallel\perp}\left( \Pi_{k\ell} T_{mn} +T_{k\ell} \Pi_{mn}\right)\right] E_{ij,\ell k}E_{ij,nm}^*
\end{align}

%%-------------------------------------------------------------------%%
%%-------------------------------------------------------------------%%
%%-------------------------------------------------------------------%%
%% Lagrange multipliers %%
%%-------------------------------------------------------------------%%
%%-------------------------------------------------------------------%%
%%-------------------------------------------------------------------%%

\section{Numerical techniques in two dimensions}\label{sctn:lagrange}
% \section{Lagrange Multipliers}\label{sctn:lagrange}
So far, we have considered a Landau theory for smectics; however, if the dynamics of the field relaxing towards a free-energy minimum are of interest, a time-dependent Ginzburg--Landau model is required 
\begin{subequations}
\begin{align}
    \label{eq:tdGL}
    \mu \frac{\partial E_{ij}}{\partial t} &= -\frac{\delta F}{\delta E_{ij}^*} + \Lambda_{ij},  
\end{align}
where $F(t) = \int d^d\vec{r} \ f\left(\vec{r},t\right)$ (\eq{eq:totalFreeEn}) is the total instantaneous free energy, $\mu$ is a mobility coefficient and $\tens{\Lambda}\left(\vec{r},t\right)$ constrains the dynamics to only those that preserve the characteristics of the smectic order parameter as described in \sctn{sctn:char}. 
In particular, as the time evolution proceeds, the tracelessness (\eq{eq:traceless}), normality (\eq{eq:normal}) and uniaxiality (\eq{eq:uni_const}) should not be allowed to numerically drift. 
The field $\tens{\Lambda}\left(\vec{r},t\right)$ is the Lagrange multiplier enforcing these conditions. 
In the following section, we explicitly determine $\tens{\Lambda}\left(\vec{r},t\right)$ for the case of $d=2$.

\subsection{Lagrange Multipliers}
To facilitate determining the Lagrange multiplier, it is beneficial to separate the elements of into their real and imaginary contributions. 
We write $\tens{E} = \tens{X} + i\tens{Y}$ for real tensors $\tens{X}$ and $\tens{Y}$ and transforms the time derivatives such that \eq{eq:tdGL} becomes 
\begin{align}
 \mu\frac{\partial X_{ij}}{\partial t} &= -\frac{\delta F}{\delta X_{ij}} +\text{Re}\left(\Lambda_{ij}\right)  \label{eq:c_g1a}\\
 \mu\frac{\partial Y_{ij}}{\partial t} &= -\frac{\delta F}{\delta Y_{ij}}+\text{Im}\left(\Lambda_{ij}\right).  \label{eq:c_g1b}
\end{align}
\end{subequations}
Consider how the conditions \eq{eq:traceless} and \eq{eq:normal} (which in two dimensions is equivalent to  \eq{eq:uni_const}) may be enforced in the numerical simulation so $\tens{E}$ maintains the desired form. 
The two conditions can be written as
\begin{subequations}
\begin{align}
     g_1(\tens{E}) &= \tr{\tens{E}} = 0 \label{eq:traceless_cond}\\
     g_2(\tens{E}) &=\det{[\tens{E},\tens{E}^*]} = 0  \label{eq:det_comut}.
\end{align}
The uniaxiality condition $\det{[\tens{E},\tens{E}^*]}$ is proportional to $\text{sin}^2(\Theta)$ for $\Theta=\text{arg}(E_{ii})-\text{arg}(E_{ij})$ for $i\ne j$.

We can rewrite $g_1$ in terms of $\tens{X}$ and $\tens{Y}$ as
\begin{align}
 g_{1a}(\tens{X})&= X_{ii} = 0\\
 g_{1b}(\tens{Y})&= Y_{ii} = 0, 
 \end{align}
which together are equivalent to the first condition (\eq{eq:traceless_cond}). 
As we are here working in two dimensions, we may use the Cayley--Hamilton Theorem to manipulate \eq{eq:det_comut} into a more numerically manageable form. 
By noting instances when $g_1=0$ occur, we can rewrite $g_2$ as
\begin{align}
\label{eq:c_g2}
    g_2(\tens{E})
    &= -\frac{1}{2}\tr{[\tens{E},\tens{E}^*]^2}
    %  &=-4\left(\text{tr}\left[YXYX-YXXY-XYYX+XYXY\right]\right)\\
    % &= -4\left(Y_{ij}X_{jk}Y_{kl}X_{li}-Y_{ij}X_{jk}X_{kl}Y_{li}-X_{ij}Y_{jk}Y_{kl}X_{li}+X_{ij}Y_{jk}X_{kl}Y_{li}\right). 
\end{align}
% \jp{If we are reducing the above to one line, I think this one is the most important. Makes it clear the use of C-H}
By evaluating the partial derivatives of each condition with respect to the elements of $\tens{X}$ and $\tens{Y}$, one can express three real Lagrange multipliers, $\lambda_{1a}$, $\lambda_{1b}$ and $\lambda_{2}$ for each of the constraints as
% \begin{align}
%  \frac{\partial g_{1a}}{\partial X_{ij}}&= \delta_{ij}\\
%  \frac{\partial g_{1a}}{\partial Y_{ij}}&= 0\\
%  \frac{\partial g_{1b}}{\partial X_{ij}}&= \delta_{ij}\\
%  \frac{\partial g_{1b}}{\partial Y_{ij}}&= 0\\
%  \frac{\partial g_{2}}{\partial X_{ij}}&= -8\left(2Y_{jk}X_{kl}Y_{li} - Y_{jk}Y_{kl}X_{li} - X_{jk}Y_{kl}Y_{li}\right)\\
%  \frac{\partial g_{2}}{\partial X_{ij}}&= -8\left(2X_{jk}Y_{kl}X_{li} - X_{jk}X_{kl}Y_{li} - Y_{jk}X_{kl}X_{li}\right).
%  \end{align}
% 
% Using these, we express each Lagrange multiplier:
\begin{align}
 \lambda_{1a}&= \frac{\delta F}{\delta X_{ii}}-\lambda_2\frac{\partial g_{2}}{\partial X_{ii}}\\
 \lambda_{1b}&= \frac{\delta F}{\delta Y_{ii}}-\lambda_2\frac{\partial g_{2}}{\partial Y_{ii}}\\
 \lambda_2&= \frac{c_1}{c_2} , 
\end{align}
where $c_1$ and $c_2$ are ungainly but straightforward
\begin{align}
 c_{1}= 
 &-\frac{\delta F}{\delta Y_{ik}}X_{kj}Y_{jp}X_{pi} -Y_{ik}\frac{\delta F}{\delta X_{kj}}Y_{jp}X_{pi} -Y_{ik}X_{kj}\frac{\delta F}{\delta Y_{jp}}X_{pi} - Y_{ik}X_{kj}Y_{jp}\frac{\delta F}{\delta X_{pi}}\nn \\
 &\ +\frac{\delta F}{\delta Y_{ik}}X_{kj}X_{jp}Y_{pi} +Y_{ik}\frac{\delta F}{\delta X_{kj}}X_{jp}Y_{pi}+Y_{ik}X_{kj}\frac{\delta F}{\delta X_{jp}}Y_{pi} +Y_{ik}X_{kj}X_{jp}\frac{\delta F}{\delta Y_{pi}}\nn \\
 &\ \  +\frac{\delta F}{\delta X_{ik}}Y_{kj}Y_{jp}X_{pi} +X_{ik}\frac{\delta F}{\delta Y_{kj}}Y_{jp}X_{pi} +X_{ik}Y_{kj}\frac{\delta F}{\delta Y_{jp}}X_{pi} +X_{ik}Y_{kj}Y_{jp}\frac{\delta F}{\delta X_{pi}}\nn \\
 &\quad -\frac{\delta F}{\delta X_{ik}}Y_{kj}X_{jp}Y_{pi} -X_{ik}\frac{\delta F}{\delta Y_{kj}}X_{jp}Y_{pi}-X_{ik}Y_{kj}\frac{\delta F}{\delta X_{jp}}Y_{pi} - X_{ik}Y_{kj}X_{jp}\frac{\delta F}{\delta Y_{pi}}
\end{align}
\begin{align}
 c_{2}= 
 &-\frac{\partial g_2}{\partial Y_{ik}}X_{kj}Y_{jp}X_{pi} -Y_{ik}\frac{\partial g_2}{\partial X_{kj}}Y_{jp}X_{pi}-Y_{ik}X_{kj}\frac{\partial g_2}{\partial Y_{jp}}X_{pi} - Y_{ik}X_{kj}Y_{jp}\frac{\partial g_2}{\partial X_{pi}}\nn \\
 &\ +\frac{\partial g_2}{\partial Y_{ik}}X_{kj}X_{jp}Y_{pi} +Y_{ik}\frac{\partial g_2}{\partial X_{kj}}X_{jp}Y_{pi} +Y_{ik}X_{kj}\frac{\partial g_2}{\partial X_{jp}}Y_{pi} +Y_{ik}X_{kj}X_{jp}\frac{\partial g_2}{\partial Y_{pi}}\nn \\
 &\ \ +\frac{\partial g_2}{\partial X_{ik}}Y_{kj}Y_{jp}X_{pi} +X_{ik}\frac{\partial g_2}{\partial Y_{kj}}Y_{jp}X_{pi} +X_{ik}Y_{kj}\frac{\partial g_2}{\partial Y_{jp}}X_{pi} +X_{ik}Y_{kj}Y_{jp}\frac{\partial g_2}{\partial X_{pi}}\nn \\
 &\quad -\frac{\partial g_2}{\partial X_{ik}}Y_{kj}X_{jp}Y_{pi} -X_{ik}\frac{\partial g_2}{\partial Y_{kj}}X_{jp}Y_{pi}-X_{ik}Y_{kj}\frac{\partial g_2}{\partial X_{jp}}Y_{pi} - X_{ik}Y_{kj}X_{jp}\frac{\partial g_2}{\partial Y_{pi}}.
\end{align}
\end{subequations}
The free energy can then be numerically minimized with respect to these constraints via the dynamics 
\begin{align}
 \frac{\partial E_{ij}}{\partial t} 
%  &= \frac{\partial X_{ij}}{\partial t} + i\frac{\partial Y_{ij}}{\partial t}\\
%  &=-\mu\bigg(\left(\frac{\delta F}{\delta X_{ij}} +i\frac{\delta F}{\delta Y_{ij}}\right) -\delta_{ij}(\lambda_{1a}+i\lambda_{1b}) - \lambda_2\left(\frac{\partial g_{2}}{\partial X_{ij}} +i\frac{\partial g_{2}}{\partial Y_{ij}}\right)\bigg)\\
 =-\frac{1}{\mu}\bigg(\frac{\delta F}{\delta E_{ij}^*} -\delta_{ij}(\lambda_{1a}+i\lambda_{1b}) - \lambda_2\frac{\partial g_{2}}{\partial E_{ij}^*}\bigg),
 \label{eq:tdGLconstrained}
\end{align}
where we have used the definition $\frac{\partial F}{\partial E_{ij}^*} = \frac{\partial F}{\partial X_{ij}} +i\frac{\partial F}{\partial Y_{ij}}$.
The time-dependent Ginzburg--Landau model given by \eq{eq:tdGLconstrained} allows the system to follow the steepest descent direction in the global free energy, while also respecting the constraints that $\tens{E}$ remain traceless and normal. 

\subsection{Extension to three dimensions}
As in two dimensions, \eq{eq:tdGL} will apply in three dimensions and the tracelessness constraint (\eq{eq:traceless}) can be dealt with in the same way as above. However, in the three dimensional case, the constraints for normality (\eq{eq:normal}) and uniaxiality (\eq{eq:uni_const}) are not equivalent to the condition det$[\tens{E},\tens{E^*}]=0$. A new form for the Lagrange multiplier $\lambda_2$ should be sought to implement the numerical relaxation described by \eq{eq:tdGL}.

%%-------------------------------------------------------------------%%
%%-------------------------------------------------------------------%%
%%-------------------------------------------------------------------%%
%% Non-dimensionalization %%
%%-------------------------------------------------------------------%%
%%-------------------------------------------------------------------%%
%%-------------------------------------------------------------------%%
\section{Non-dimensionalization}\label{sctn:nondim}
The Landau theory has been presented above in a form with material coefficients. However, it is informative to non-dimensionalize the free energy density, which reveals the characteristic material length scales of the system. 
Continuing to neglect surfaces or external fields, and non-dimensionalizing the free energy density by the scale of $A$ gives 
\begin{align}
 \tilde{f} &= \frac{\tilde{A}}{2}E_{ij}E_{ji}^* + \frac{\tilde{C}}{4}\left(E_{ij}E_{ij}^*\right)^2 + \xi^2 \left\{ \left[ \Pi_{k\ell}+\frac{b_1^\perp}{b_1^\parallel} T_{k\ell} \right] E_{ij,k} E_{ij,\ell}^*     \right. \nn\\
       &\qquad \left. + \lambda^2 \left[ T_{k\ell} T_{mn} + \frac{b_2^\parallel}{b_2^\perp} \Pi_{k\ell} \Pi_{mn} + \frac{b_2^{\parallel\perp}}{b_2^\perp} \left(\Pi_{k\ell} T_{mn} + T_{k\ell} \Pi_{mn}\right)\right] E_{ij,\ell k} E_{ij,nm}^* \right\}, 
\end{align}
where $\tilde{C}=C/\abs{A}$ and $\tilde{A}$ is the non-dimensional distance from the transition point, which can take the values $\tilde{A}=+1$ for the isotropic phase, $\tilde{A}=-1$ for the lamellar phase and $\tilde{A}=0$ at the transition. 
The non-dimensionalization reveals five material length scales, in addition to the lamellar wave length $\thick=2\pi/q_0$. 
The lengthscale 
\begin{align}
    \label{eq:persistenceLength}
    \xi &= \sqrt{ \frac{b_1^\parallel}{\abs{A}}}
\end{align} 
is the lamellar in-plane coherence length and the penetration depth is
\begin{align}
    \label{eq:penetrationDepth}
     \lambda &= \sqrt{ \frac{b_2^\perp}{b_1^\parallel} } .
\end{align} 
A second coherence length is $\xi^\perp=\xi \sqrt{ b_1^\perp/b_1^\parallel} = \sqrt{ b_1^\perp /A}$ and the two additional penetration depths are $\lambda^\parallel = \sqrt{ b_2^{\parallel} / b_1^\parallel}$ and $ \lambda^{\parallel\perp} =\sqrt{b_2^{\parallel\perp} / b_1^\parallel}$. 
The order of magnitude of these lengths can be approximated from experimental values. 
Smectic layer thickness is approximately $d_0\simeq3.5\text{nm}$~\cite{deVries1977}, the penetration depth is approximately $\lambda\simeq2\text{nm}$~\cite{chen2000} and the coherence length $\xi$ can be somewhat larger ($\simeq20\text{nm}$~\cite{chu1977}) or smaller ($\simeq0.5\text{nm}$~\cite{Zappone2020}).
The ratio of coherence lengths $\xi/\xi^\perp\simeq1-10$~\cite{chu1977,lubensky1983,ambrozic2004}, but the ratios of penetration depths remain relatively unexplored.
Liquid crystaline {\em sm-A} or {\em sm-C}, would also have three Frank coefficients (as in \eq{eq:nemDef}), producing three additional length scales, for a total number of eight physical length scales --- the number of characteristic length scales expected for a smectic material~\cite{oswald2005}. 

Furthermore, we identify the Ginzburg parameter~\cite{renn1988,Lubensky1990,Zappone2020}
\begin{align}
    \boxed{ \kappa = \frac{\lambda}{\xi} 
                    = \frac{ \left(\abs{A} b_2^\perp\right)^{1/2} }{ b_1^\parallel } } \ .
\end{align}
In superconductors, $\kappa<1/\sqrt{2}$ is a type-I systems, while $\kappa>1/\sqrt{2}$ is a type-II~\cite{degennes1972}. 
Our $\tens{E}$-based theory is able to model both regimes. 

To write the non-dimensionalized free energy density more cleanly, we non-dimensionalize units of length by the persistence length $\xi$ from \eq{eq:persistenceLength} and define dimensionless elasticity modulii $\tilde{b}_1^\perp=b_1^\perp/b_1^\parallel$, $\tilde{b}_2^\parallel=b_2^\parallel/b_2^\perp$ and $\tilde{b}_2^{\parallel\perp}=b_2^{\parallel\perp}/b_2^\perp$. 
The non-dimensionalized free energy density is then written in terms of the Ginzburg parameter
% \begin{align}
% % \begin{empheq}[boxtype=\fbox]
%     \tilde{f} &= \frac{\tilde{A}}{2}E_{ij}E_{ji}^* + \frac{\tilde{C}}{4}\left(E_{ij}E_{ij}^*\right)^2 + \left[  \Pi_{k\ell}+ \tilde{b}_1^\perp T_{k\ell} \right] E_{ij,\tilde{k}} E_{ij,\tilde{\ell}}^* 
%      \nn\\ &
%      + \kappa^2 \left[  T_{k\ell} T_{mn} + \tilde{b}_2^\parallel \Pi_{k\ell} \Pi_{mn}  + \tilde{b}_2^{\parallel\perp} \left(\Pi_{k\ell} T_{mn} + T_{k\ell} \Pi_{mn}\right)\right] E_{ij,\tilde{\ell}\tilde{k}} E_{ij,\tilde{n}\tilde{m}}^* \ , 
%      \label{eq:nondim}
% % \end{empheq}
% \end{align} 

\begin{equation}
	\label{eq:nondim}
    \boxed{
    	\tilde{f} = \frac{\tilde{A}}{2}E_{ij}E_{ji}^* + \frac{\tilde{C}}{4}\left(E_{ij}E_{ij}^*\right)^2 + \left[  \Pi_{k\ell}+ \tilde{b}_1^\perp T_{k\ell} \right] E_{ij,\tilde{k}} E_{ij,\tilde{\ell}}^* 
     + \kappa^2 \left[  T_{k\ell} T_{mn} + \tilde{b}_2^\parallel \Pi_{k\ell} \Pi_{mn}  + \tilde{b}_2^{\parallel\perp} \left(\Pi_{k\ell} T_{mn} + T_{k\ell} \Pi_{mn}\right)\right] E_{ij,\tilde{\ell}\tilde{k}} E_{ij,\tilde{n}\tilde{m}}^* \ , 
    }
\end{equation}
where tildes over the indices indicate that the gradients have been non-dimensionalized by the coherence length, $\xi$. 

It is also worth noting that non-dimensionalizing the surface free energy per unit area produces another set of length scales. In the case of the uniform Rapini--Papoular surface anchoring, non-dimensionalization by the layer compression elastic modulus $b_1^\parallel$, produces the de~Gennes--Kleman extrapolation length $b_1^\parallel/W_0$. 
However, each of the anchoring strengths produce a length scale $b_1^\parallel/W_i$ for $i=0,1,2$ in \eq{eq:anchRP} or \eq{eq:anchDeg}.

%%-------------------------------------------------------------------%%
%%-------------------------------------------------------------------%%
%%-------------------------------------------------------------------%%
%% Equilibrium %%
%%-------------------------------------------------------------------%%
%%-------------------------------------------------------------------%%
%%-------------------------------------------------------------------%%
\section{Equilibrium}\label{sctn:eq}

We now consider the equilibrium values that arise from this phenomenological Landau theory. 
To make these connections more explicit, we return to the dimensional form of the theory (\eq{eq:total}) and consider the free energy density when \textbf{not} subjected to any deformations, \ie such that $\tens{E}$ is constant. 
To determine microscale properties, such as the layer thickness, we must consider the microscopic variation across layers. 
To explore microscopic properties of well-aligned layers, we must apply covariant derivatives. 

%%-------------------------------------------------------------------%%
%%-------------------------------------------------------------------%%
%%-------------------------------------------------------------------%%
%% Covariance %%
%%-------------------------------------------------------------------%%
%%-------------------------------------------------------------------%%
%%-------------------------------------------------------------------%%
\subsection{Covariance}\label{sctn:covar}

Basic gradients has been employed in \eq{eq:comp} and \eq{eq:curv}, rather than covariant derivatives~\cite{degennes1972,chen1976,renn1988,renn1992,lukyanchuk1998,park2006,joo2007,Calderer2008}. 
In smectic theories, it is common and often necessary to employ the covariant derivative  
\begin{align}
  \label{eq:covar}
  \vec{D} \equiv \del -iq_0\vec{N} .
\end{align}
The present model does not require a covariant derivative because $\tens{E}$ is a hydrodynamic variable that describes the macroscopic configuration and does not account for microscopic variation. 
If instead we had chosen to write the free energy in terms of an explicit, local, plane-wave-based tensor $\tens{\mathcal{E}} = \tens{E} e^{iq_0\vec{N}\cdot\vec{r}}$ (\eq{eq:microE}), then the situation would require covariant derivative with respect to the metric of the microscale surfaces defining the smectic layers. 
We now show that the free energy densities expressed in the previous sections are equivalent to the microscopic form
\begin{subequations}
\begin{align}
	f^\text{bulk} &= \frac{A}{2}\mathcal{E}_{ij}\mathcal{E}_{ji}^* + \frac{C}{4}\left(\mathcal{E}_{ij}\mathcal{E}_{ij}^*\right)^2+\ldots 
	\label{eq:bulkMicro}
	\\
	f^\text{comp} &= b_1^\parallel \Pi_{k\ell} D_{k}\mathcal{E}_{ij} \ D_{\ell}^*\mathcal{E}_{ij}^* + b_1^\perp T_{k\ell} D_{k}\mathcal{E}_{ij} \ D_{\ell}^*\mathcal{E}_{ij}^* 
	\label{eq:compMicro}
	\\
	f^\text{curv} &= b_2^\parallel \Pi_{k\ell}D_lD_k\mathcal{E}_{ij} \Pi_{mn}D_n^*D_m^*\mathcal{E}_{ij}^* + b_2^\perp T_{k\ell}D_lD_k\mathcal{E}_{ij} T_{mn}D_n^*D_m^*\mathcal{E}_{ij}^* \nn \\
	&\qquad + b_2^{\parallel\perp} \left(\Pi_{k\ell}D_lD_k\mathcal{E}_{ij} T_{mn}D_n^*D_m^*\mathcal{E}_{ij}^*  + T_{k\ell}D_lD_k\mathcal{E}_{ij} \Pi_{mn}D_n^*D_m^*\mathcal{E}_{ij}^*\right). 
	\label{eq:curvMicro}
\end{align}
\end{subequations}
Demonstrating the equivalence between these descriptions does not require every term in the free energies be considered explicitly. 
Instead, it is sufficient to note
\begin{enumerate}
	\item The bulk free energy is unchanged because 
	\begin{align}
		\mathcal{E}_{ij}\mathcal{E}_{ji}^* &= E_{ij}e^{iq_0\vec{N}\cdot\vec{r}}E_{ji}^*e^{-iq_0\vec{N}\cdot\vec{r}} = E_{ij}E_{ji}^* .
	\end{align}
	\item Denoting the covariant derivative by a semicolon $;$, the compression and curvature free energies are unchanged because 
	\begin{align}
		\mathcal{E}_{ij;k} \mathcal{E}_{ij;k}^* 
		    &= D_k\mathcal{E}_{ij} D_\ell^*\mathcal{E}_{ij}^* 
		    = \left[ \left(\partial_k - iq_0N_k\right)E_{ij} e^{iq_0 \vec{N}\cdot\vec{r}} \right] \left[ \left(\partial_\ell + iq_0N_\ell\right)E_{ij}^* e^{-iq_0 \vec{N}\cdot\vec{r}} \right] \nn\\
		  %  &=  \left[ e^{iq_0 \vec{N}\cdot\vec{r}} \partial_kE_{ij}  + E_{ij} \partial_k e^{iq_0 \vec{N}\cdot\vec{r}} - iq_0N_k E_{ij} e^{iq_0 \vec{N}\cdot\vec{r}}  \right]   \left[ e^{-iq_0 \vec{N}\cdot\vec{r}} \partial_\ell E_{ij}^* + E_{ij}^* \partial_\ell e^{-iq_0 \vec{N}\cdot\vec{r}} + iq_0N_\ell E_{ij}^* e^{-iq_0 \vec{N}\cdot\vec{r}} \right] \nn\\
		  %   &= \left(\partial_kE_{ij}\right)e^{iq_0 \vec{N}\cdot\vec{r}}\left(\partial_\ell E_{ij}^*\right)e^{-iq_0 \vec{N}\cdot\vec{r}} 
		     &= E_{ij,k} E_{ij,\ell}^*. 
	\end{align}
\end{enumerate}
Thus, the total free energy is unchanged, which is not surprising since this is expressly the function of a {\em covariant} derivative. 
Since the two forms are thus equivalent, we might question the difference in working with $\tens{{E}}$ compared to $\tens{\mathcal{E}}$.
Employing $\tens{\mathcal{E}}$ introduces covariant derivatives, depending explicitly on the layer normal direction and this would necessitate a numerical scheme that diagonalizes $\tens{\mathcal{E}}$ at every time step and uses the instantaneous $\vec{N}$. 
For this reason, evolving $\tens{E}$ is not only more numerically straightforward but is also more elegant in that the dynamics involves only $\tens{E}$ itself without explicit reference to its eigenvalues or vectors; solving for $\vec{N}$ and $\psi$ is a post-simulation analysis only performed at the times of interest.
Therefore, we prefer to work with $\tens{E}$. 
However, considering $\tens{\mathcal{E}}$ is useful for determining the microscale, equilibrium properties of the lamellar phase, such as wave number $q_0$ (see \sctn{sctn:equilibrium}). 

%%-------------------------------------------------------------------%%
%%-------------------------------------------------------------------%%
%%-------------------------------------------------------------------%%
%% Equilibrium %%
%%-------------------------------------------------------------------%%
%%-------------------------------------------------------------------%%
%%-------------------------------------------------------------------%%
\subsection{Equilibrium Values}\label{sctn:equilibrium}
To determine the equilibrium values, we consider the free energy density when the smectic is {\em not subjected to any deformations}. 
That is to say, $\tens{E}$ is constant at equilibrium. 
Holding $\tens{E}$ constant in \eq{eq:total} gives
\begin{align}
    \label{eq:undeformed_grad}
    f^\text{eq} = \frac{A}{2}\varrho \app + \frac{C}{4}\varrho^2 \ps^4 .
\end{align}
From \eq{eq:undeformed_grad}, we see that $f^\text{eq}$ is independent of the constant $\phi$ and $\vec{N}$ values. 
From \eq{eq:undeformed_grad}, the equilibrium value for the order parameter $\ps$ can be found to be
% For the amplitude,
\begin{align}
\label{eq:undeformed_grad_2}
 \frac{\partial f^\text{eq}}{\partial \ps} &= 0 = A\varrho \ps + C\varrho^2 \ps^3  \nn\\
 \ps^\text{eq} &= \sqrt{-\frac{A}{C\varrho}} .
\end{align}
When $A<0$, the material is in the isotropic phase, while $A>0$ is the lamellar phase for $C>0$. 
While the macroscopic equilibrium degree of ordering $\ps^\text{eq}$ can be found from the \eq{eq:undeformed_grad_2}, microscopic equilibrium properties cannot. 
To estimate microscopic equilibrium properties such as the wave number $q_0$, the covariance must be employed. 
\sctn{sctn:covar} demonstrated that applying $\del$ to $\tens{E}$, or applying the covariant derivative $\vec{D}=\del-iq_0\vec{k}$ to $\tens{\mathcal{E}}$ are equivalent and do not introduce a dependence on the wave number $q_0$. 
To explore corrections resulting from the explicit inclusion of the wave number, we instead apply the covariant derivative to our macroscale $\tens{E}$.

The bulk free energy is unchanged from \eq{eq:bulkTrad} and \eq{eq:undeformed_grad} so we consider $f^\text{comp}$ and $f^\text{curv}$ for constant $\tens{E}$. 
We consider the parallel, perpendicular and mixed terms in turn. 
The compression free energy (\eq{eq:compMicro}) can be split into $f^\text{comp} = f^\text{comp}_\parallel + f^\text{comp}_\perp$. 
The parallel and perpendicular contributions for constant $\vec{N}$ and $\psi$ become
\begin{subequations}
\begin{align}
%  f^\text{comp}_\parallel &= b_1^\parallel \Pi_{k\ell} D_k{\mathcal{E}}_{ij} D_\ell^* {\mathcal{E}}_{ij}^*
 f^\text{comp}_\parallel &= b_1^\parallel \Pi_{k\ell} D_k{{E}}_{ij} D_\ell^* {{E}}_{ij}^*
                        %  = b_1^\parallel N_{k}N_{\ell} \left(\nabla_k -iq_0N_k\right){\mathcal{E}}_{ij} \left(\nabla_k + iq_0N_\ell\right) {\mathcal{E}}_{ij}^* \\
                        %  &= b_1^\parallel N_{k}N_{\ell} \left(-iq_0N_k\right){\mathcal{E}}_{ij} \left(iq_0N_\ell \right) {\mathcal{E}}_{ij}^* 
                        %  = b_1^\parallel q_0^2 {\mathcal{E}}_{ij} {\mathcal{E}}_{ij}^* \\
                         = b_1^\parallel q_0^2 \varrho \app \\
%  f^\text{comp}_\perp &= b_1^\perp T_{k\ell} D_k {\mathcal{E}}_{ij} D_\ell^* {\mathcal{E}}_{ij}^*
  f^\text{comp}_\perp &= b_1^\perp T_{k\ell} D_k {{E}}_{ij} D_\ell^* {{E}}_{ij}^*
                    %  = b_1^\perp \left(\delta_{k\ell} - N_kN_\ell \right) \left(\nabla_k -iq_0N_k\right) {\mathcal{E}}_{ij} \left(\nabla_k + iq_0N_\ell\right) {\mathcal{E}}_{ij}^* \\
                    %  &= b_1^\perp q_0^2 \left(\delta_{k\ell} - N_kN_\ell \right) N_k {\mathcal{E}}_{ij} N_\ell {\mathcal{E}}_{ij}^* \\
                    %  &= b_1^\perp q_0^2 \left(N_k {\mathcal{E}}_{ij} N_k {\mathcal{E}}_{ij}^* - N_kN_\ell N_k {\mathcal{E}}_{ij} N_\ell {\mathcal{E}}_{ij}^* \right) \\
                     = 0.
\end{align}
Likewise, we consider the curvature-type free energy density contributions. 
For convenience, we define $\Delta_\parallel \equiv \Pi_{k\ell} D_\ell D_k^* = \left[ (N_k\nabla_k)^2 +q_0^2 \right]$ and $\Delta_\perp \equiv T_{k\ell} D_\ell D_k^* = \left[ \nabla_k\nabla_k - (N_k\nabla_k)^2 \right]$. 
Thus, the free energy contributions become
\begin{align}
%  f^\text{curv}_\parallel &= b_2^\parallel \Delta_\parallel \mathcal{E}_{ij} \Delta_\parallel \mathcal{E}_{ij}^* 
 f^\text{curv}_\parallel &= b_2^\parallel \Delta_\parallel {E}_{ij} \Delta_\parallel {E}_{ij}^* 
                        %  = b_2^\parallel \left[\left(N_k\nabla_k\right)^2 + q_0^2\right] \mathcal{E}_{ij} \left[\left(N_\ell\nabla_\ell\right)^2 + q_0^2\right] \mathcal{E}_{ij}^* 
                        %  = b_2^\parallel q_0^4 \mathcal{E}_{ij}\mathcal{E}_{ij}^* \nn \\ 
                         = b_2^\parallel q_0^4 \varrho \app\\
%  f^\text{curv}_\perp &= b_2^\perp \Delta_\perp \mathcal{E}_{ij} \Delta_\perp \mathcal{E}_{ij}^*
 f^\text{curv}_\perp &= b_2^\perp \Delta_\perp {E}_{ij} \Delta_\perp {E}_{ij}^*
                    %  = b_2^\perp \left[\nabla^2 - \left(N_k\nabla_k\right)^2\right] \mathcal{E}_{ij} \left[\nabla^2 - \left(N_k\nabla_k\right)^2\right] \mathcal{E}_{ij}^* \nn\\
                     = 0 \\
%  f^\text{curv}_{\parallel\perp} &= b_2^{\parallel\perp} \left(\Delta_\parallel \mathcal{E}_{ij} \Delta_\perp \mathcal{E}_{ij}^* +\Delta_\perp \mathcal{E}_{ij} \Delta_\parallel \mathcal{E}_{ij}^* \right) 
 f^\text{curv}_{\parallel\perp} &= b_2^{\parallel\perp} \left(\Delta_\parallel {E}_{ij} \Delta_\perp {E}_{ij}^* +\Delta_\perp {E}_{ij} \Delta_\parallel {E}_{ij}^* \right) 
                            %   &=b_2^{\parallel\perp} \left(\Delta_\parallel \mathcal{E}_{ij} \left[\nabla^2 - \left(N_k\nabla_k\right)^2\right] \mathcal{E}_{ij}^* +  \left[\nabla^2 - \left(N_k\nabla_k\right)^2\right] \mathcal{E}_{ij}  \Delta_\parallel\mathcal{E}_{ij}^*\right) \nn\\
                            = 0.
\end{align}
\end{subequations}
Summing these, the total microscopic free energy density for an undeformed smectic is
\begin{align}
    \label{eq:undeformed}
    f^\text{eq} = \frac{A}{2}\varrho \app + \frac{C}{4}\varrho^2 \ps^4 + b_1^\parallel q_0^2 \varrho \app + b_2^\parallel q_0^4 \varrho \app .
\end{align}
From \eq{eq:undeformed}, the equilibrium values for the modulus $\ps$ and the wave number $q_0$ can be determined. 
For the amplitude,
\begin{align}
 \frac{\partial f^\text{eq}}{\partial \ps} &= 0 = A\varrho \ps + C\varrho^2 \ps^3 + 2b_1^\parallel q_0^2 \varrho \ps + 2b_2^\parallel q_0^4 \varrho \ps \nn\\
 \ps^\text{eq} &= \sqrt{-\frac{A'}{C\varrho}} ,
\end{align}
where $A' \equiv A + 2b_1^\parallel q_0^2 + 2b_2^\parallel q_0^4$.
The wave number $q_0$ is assumed to be the equilibrium value since $\phi$ takes into account displacements of the layers and merely specifies the overlayed plane wave. 
We find this equilibrium wave number to be
\begin{align}
 \frac{\partial f^\text{eq}}{\partial q_0} &= 0 = 2b_1^\parallel q_0 \varrho \app + 4b_2^\parallel q_0^3 \varrho \app \nn\\
  q_0^2 &= - \frac{b_1^\parallel}{2b_2^\parallel} .
\end{align}
These equilibrium values depend only on $b_1^\parallel$ and $b_2^\parallel$, which are the parallel-term elastic constants. 
The lamellar wave number in units of coherence length is
\begin{align}
 \tilde{q}_0^2 &= \left(\xi q_0\right)^2 = -\frac{b_1^{\parallel \ 2}}{2\abs{A}b_2^\parallel} . 
 \label{eq:wavenum}
\end{align}
By taking $b_1^\parallel<0$ and $b_2^\parallel>0$ successfully defines the microsopic wave number in terms of elastic constants, in agreement with scalar theories~\cite{das2008,mukherjee2013}. 
However on physical grounds and following previous simulations~\cite{abukhdeir2008, Abukhdeir2009, Abukhdeir2010}, we expect deformations to come at a free energy cost, which suggests $b_i^\parallel>0$ for $i\in\{1,2\}$. 
Therefore, we conclude that this phenomenological Landau theory leaves $q_0$ non-determined. 
We expect that in any future theory that extends this work to {\em sm-C}, the wave number will be expressed in terms of both $b_1^\parallel$, $b_2^\parallel$ and coefficients of coupling terms between $\tens{E}$ and the nematic tensor $\tens{Q}$~\cite{mukherjee2007,Mukherjee2004,Mukherjee2021a}.

%%-------------------------------------------------------------------%%
%%-------------------------------------------------------------------%%
%%-------------------------------------------------------------------%%
%% First simplification %%
%%-------------------------------------------------------------------%%
%%-------------------------------------------------------------------%%
%%-------------------------------------------------------------------%%
\section{Simplifying assumptions}
\label{sctn:reducing}
At first glance, the distortion free energy densities (\eq{eq:comp} and \eq{eq:curv}) appear unwieldy. 
However, as in $\tens{Q}$-based Landau--de~Gennes theory, we can reduce complications substantially by making the simplifying choice of a one-elastic-constant approximation. 
Under what circumstances does the complex-tensor-based Landau theory for smectics reduce to simpler forms? 

\subsection{One-constant approximation}\label{sctn:oneconst}
We begin by considering a one-constant approximation for both the compression free energy (\eq{eq:comp}) and the curvature free energy (\eq{eq:curv}). 
The definition of the perpendicular projection operator $\tens{T} = \tens{\delta} - \tens{\Pi}$ (\eq{eq:projPerp}) can be inserted into \eq{eq:comp}, such that all terms involving the parallel projection operator can be grouped
\begin{align}
 f^\text{comp} 
%                &= b_1^\parallel \Pi_{k\ell} E_{ij;k} E_{ij;\ell}^* + b_1^\perp T_{k\ell} E_{ij;k} E_{ij;\ell}^*  \\
               &= b_1^\perp E_{ij,k} E_{ij,k}^* + \left(b_1^\parallel - b_1^\perp\right) \Pi_{k\ell} E_{ij,k} E_{ij,\ell}^* . 
\end{align}
In this form, it becomes clear that a one-constant approximation
\begin{align}
    \label{eq:oneConst_b1}
	b_1^\parallel &= b_1^\perp \equiv b_1 ,
\end{align}
removes the need for projection operators and simplifies the form of the free energy to $f^\text{comp} = b_1 E_{ij,k} E_{ij,k}^*$. 
\begin{figure*}[tb]
    \centering
    \includegraphics[width=0.75\textwidth]{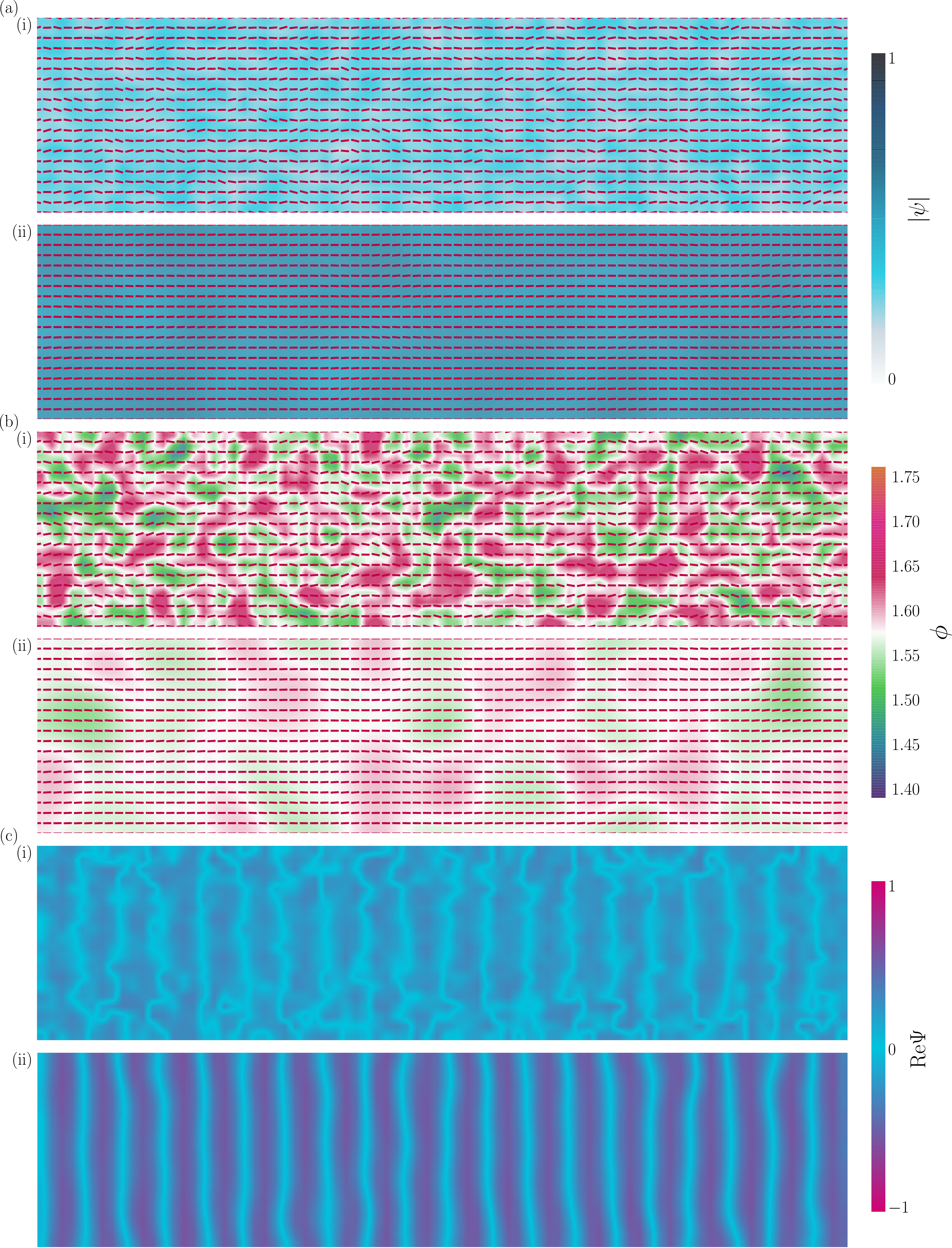}
    \caption{\small
        Numerical simulation of perturbed layers in a periodic domain using the one constant approximation. 
        Here, $\Delta x =\xi$ and $\Delta t =0.001\mu$. 
        The physical parameters are $A=-1$, $C=2$ and $\kappa^2=0.6$. 
        Initial conditions are such that, at time $t=0$, $\ps$ is uniformly distributed in the region $(0.2, 0.4)$; $\phi\in(\frac{\pi}{2}-0.15, \frac{\pi}{2}+0.15)$ and $\vec{N}=(\cos(\theta),\sin(\theta))$ for $\theta\in(-0.5,0.5)$.  
        We observe a relaxation towards a more ordered state at times (i) $t=0.025\mu$ and (ii) $t=0.875\mu$. 
        \textbf{(a)} The extent of the layering $\ps\left(\vec{r},t\right)$. 
        The red bars correspond to the layer normal, $\vec{N}\left(\vec{r},t\right)$. 
        At later times $\ps$ relaxes to the equilibrium value ($\ps_{eq}=1$).
        \textbf{(b)}Plot showing the phase $\phi\left(\vec{r},t\right)$.  
        The red bars correspond to the layer normal, $\vec{N}$. 
        Over time gradients of $\phi$ fade, demonstrating a reduction in layer compression/dilation.
        \textbf{(c)}Plot showing the layer visualisation $\RE{\Psi}$ for $q_0=0.913$.
    }
    \label{fig:sim}
\end{figure*}

The same can be done for the curvature contribution (\eq{eq:curv})
\begin{align}
 f^\text{curv} &= \left(b_2^\parallel + b_2^\perp - 2b_2^{\perp\parallel}\right)\Pi_{kl}E_{ij,kl}\Pi_{mn}E_{ij,mn}^* + b_2^\perp E_{ij,kk}E_{ij,mm}^* \nn\\
 &\qquad + \left(b_2^{\perp\parallel}-b_2^\perp\right) \left( E_{ij,kk}\Pi_{mn}E_{ij,mn}^* + \Pi_{kl}E_{ij,kl}E_{ij,mm}^* \right) . 
\end{align}
Once again, we can choose a one-constant approximation
\begin{align}
    \label{eq:oneConst_b2}
	b_2^\parallel &= b_2^\perp =b_2^{\perp\parallel}\equiv b_2 ,
\end{align}
which again eliminates all the projection operators and simplifies this free energy term to $f^\text{curv} = b_2 E_{ij,kk} E_{ij,mm}^*$. 
This pair of one-elastic constant approximations, thus produces a non-dimensionalized free energy density which can be written in terms of the Ginzburg parameter as simply
\begin{align}
    \label{eq:oneConstFE}
    \boxed{ \tilde{f} = \frac{\tilde{A}}{2}E_{ij}E_{ji}^* + \frac{\tilde{C}}{4}\left(E_{ij}E_{ij}^*\right)^2 + E_{ij,\tilde{k}} E_{ij,\tilde{k}}^* 
     + \kappa^2  E_{ij,\tilde{k}\tilde{k}}E_{ij,\tilde{m}\tilde{m}} ^* } \ . 
\end{align}
This shows that the general Landau theory for a complex-tensor order parameter reduces to the previously used form~\cite{paget2022}. 

{Substituting the one-elastic constant free energy (\eq{eq:oneConstFE}) in the time-dependent Ginzburg--Landau model (\eq{eq:tdGLconstrained}) and numerically integrating $\tens{E}$ via a two-step Adams–Bashforth method shows the relaxation dynamics (\fig{fig:sim}). 
This simulation is performed in a thin periodic domain. At time $t=0$, we set $\ps$ uniformly distributed in the region $(0.2, 0.4)$; $\phi\in(\frac{\pi}{2}-0.15, \frac{\pi}{2}+0.15)$ and $\vec{N}=(\cos(\theta),\sin(\theta))$ for $\theta\in(-0.5,0.5)$ and construct $\tens{E}\left(\vec{r},t\right)$ using these. 
In simulations, $\tens{E}\left(\vec{r},t\right)$ is calculated with $\psi$ and $\vec{N}$ found \textit{ex post facto} --- $\ps$ found via \eq{eq:calcModulus}, $\phi$ via \eq{eq:calcPhase} and $\vec{N}$ via eigen-decomposition of the real tensor $e^{-i\phi}\tens{E}$. }

At early times ($t\ll\mu$), the degree of ordering $\ps\left(\vec{r},t\right)$ is small and the layer normal $\vec{N}\left(\vec{r},t\right)$ is isotropically disordered (\fig{fig:sim}; top). 
Similarly, the phase $\phi$ is noisy (\fig{fig:sim}; middle). 
Over time, the gradients in $\phi$ decrease, showing a reduction in local dilation and compression (\movie{1}). 
Alongside this, the scalar order parameter $\ps$, denoting the extent of the layering, relaxes towards its equilibrium value (\movie{2}).  
Both of these aspects can be observed clearly in plots of $\RE{\Psi}$, which depicts the layering. 
As the time approaches the relaxation time $\mu$, the layers become more uniform and well defined, which can also be seen in the supplementary movies (\movie{3}). 

This numerical scheme has been applied in simulations for \fig{fig:sim} and in Ref~\cite{paget2022}. 
In both cases, the one-constant approximation has been made for the free energy density $f(\vec{r},t)$ (\eq{eq:oneConstFE}). 
The complication for anisotropic constants is that the projection operators must vary with $\tens{E}$ (\eq{eq:PI_def}) and must be included in the variation of the free energy in the Ginzburg--Landau equation (\eq{eq:tdGL}). 
However, the numerical approach is expected to be equally applicable in this more complicated situation.

%%-------------------------------------------------------------------%%
%%-------------------------------------------------------------------%%
%%-------------------------------------------------------------------%%
%% Comparison to other models %%
%%-------------------------------------------------------------------%%
%%-------------------------------------------------------------------%%
%%-------------------------------------------------------------------%%

\subsection{Comparison to alternative descriptions}\label{sctn:review}
Determining the equilibrium values of the smectic in \sctn{sctn:equilibrium} required consideration of the free energy when $\tens{E}$ was entirely fixed. 
In this section, we allow one aspect of $\tens{E}$ to vary while holding the others fixed. 
This enables us to compare $\tens{E}$-theory to models of smectics that solely treat one aspect or another. 

\subsubsection{Elastic de~Gennes--McMillan theory}
We begin by presuming that the layer normal $\vec{N}$ is fixed globally along a constant axis and that the material is sufficiently deep within the lamellar phase such that $\ps$ is constant. 
In this case, only compression deformation is permitted and only $\phi\left(\vec{r},t\right)$ can vary. 
To determine the free energy under these conditions, we substitute \eq{eq:tensE} into the compression and curvature free energies to find
\begin{subequations}
\begin{align}
    \frac{f^\text{comp}}{\varrho\ps^2} 
        &= b_1^\perp\del\phi\cdot\del\phi + \left(b_1^\parallel - b_1^\perp\right)\left(\vec{N}\cdot\del\phi\right)^2 \\
    \frac{f^\text{curv}}{\varrho\ps^2} 
        &= b_2^\perp \left( \left(\del\phi\cdot\del\phi\right)^2+ \left(\nabla^2\phi\right)^2\right) 
        + \left(b_2^\perp + b_2^\parallel -2b_2^{\parallel\perp}\right)\left( \left(\vec{N}\cdot\del\phi\right)^4 + \left((\vec{N}\cdot\del)^2\phi\right)^2\right) \nn\\
        &\qquad + 2\left(b_2^{\parallel\perp}-b_2^\perp\right)\left( \left(\del\phi\cdot\del\phi\right)\left(\vec{N}\cdot\del\phi\right)^2  + (\vec{N}\cdot\del)^2\phi \nabla^2\phi\right).
\end{align}
Since $\vec{N}$ is held fixed and the compression can only occur in the direction of the layer normal, all the gradients $\del\phi$ must be in the layer normal direction. 
Thus, $\del \to \vec{N}\cdot\del = \nabla_\parallel$ in this case. However, the Laplacian can have other contributions and, in fact, one would expect that the simplest deformations would come from curvatures perpendicular to the layers $\nabla_\perp^2 =\nabla^2-\nabla_\parallel^2$. 
Substituting these definitions into the free energies gives
\begin{align}
    \frac{f^\text{comp}}{\varrho\ps^2} 
        % &= b_1^\perp\left(\nabla_\parallel\phi\right)^2 + \left(b_1^\parallel - b_1^\perp\right)\left(\nabla_\parallel\phi\right)^2 \nn\\
        &= b_1^\parallel\left(\nabla_\parallel\phi\right)^2\\
    \frac{f^\text{curv}}{\varrho\ps^2} 
    % &= b_2^\perp \left[ \left(\nabla_\parallel\phi\right)^2+ \left(\nabla^2\phi\right)^2\right] 
    %     + \left(b_2^\perp + b_2^\parallel -2b_2^{\parallel\perp}\right)\left[ \left(\nabla_\parallel\phi\right)^4 + \left(\nabla_\parallel^2\phi\right)^2\right] \nn\\
    %     &\qquad + 2\left(b_2^{\parallel\perp}-b_2^\perp\right)\left[ \left(\nabla_\parallel\phi\right)^2\left(\nabla_\parallel\phi\right)^2  + \nabla_\parallel^2\phi \nabla^2\phi\right] \nn\\
    %  &= b_2^\perp \left(\nabla_\parallel\phi\right)^2 + \left(b_2^\parallel-b_2^\perp\right)\left(\nabla_\parallel\phi\right)^4 + b_2^\perp \left(\nabla^2\phi\right)^2
    %     + \left(b_2^\perp + b_2^\parallel -2b_2^{\parallel\perp}\right) \left(\nabla_\parallel^2\phi\right)^2 + 2\left(b_2^{\parallel\perp}-b_2^\perp\right) \nabla_\parallel^2\phi \nabla^2\phi \nn \\
    % &= b_2^\perp  \left[1 + \left(\frac{b_2^\parallel}{b_2^\perp}-1\right)\left(\nabla_\parallel\phi\right)^2\right] \left(\nabla_\parallel\phi\right)^2 \nn\\
    %     &\qquad + b_2^{\parallel\perp} \left(\nabla^2\phi\right)^2
    %         + \left(b_2^\parallel - b_2^{\parallel\perp}\right) \left(\nabla_\parallel^2\phi\right)^2 + \left(b_2^\perp-b_2^{\parallel\perp}\right) \left(\nabla_\perp^2\phi\right)^2 .
    &= b_2^\parallel  \left(\nabla_\parallel\phi\right)^2 
         + b_2^{\parallel\perp} \left(\nabla^2\phi\right)^2+ \left(b_2^\parallel - b_2^{\parallel\perp}\right) \left(\nabla_\parallel^2\phi\right)^2 + \left(b_2^\perp-b_2^{\parallel\perp}\right) \left(\nabla_\perp^2\phi\right)^2 .
\end{align}
\end{subequations}
The compressional term $f^\text{comp}$ has a direct physical interpretation --- by assuming that all gradients are in the layer normal direction, the only elasticity that matters is the layer compression elastic modulus $b_1^\parallel$. 
However, this is then supplemented by the fact that curvature-type deformations $f^\text{curv}$ might necessitate layer displacements such that $b_2^\perp$ (with a higher order correction~\cite{Walker2010}) contributes to this type of deformation. 
The curvature contribution to the deformation free energy density can be simplified to directly correspond to physically transparent descriptions~\cite{degennes1993}. 
To do so, we make two assumptions: 
(i) The higher order term involving $\left(\nabla_\parallel\phi\right)^2$ can be neglected, and
(ii) the elastic constants, $b_2^\parallel$, $b_2^\perp$ and $b_2^{\parallel\perp}$, are simply related. 
To make direct comparison to other models, we must make choices relating the elastic constants $b_2^\parallel$, $b_2^\perp$ and $b_2^{\parallel\perp}$. 

\paragraph{One-constant approximation}
If we continue with the one-constant approximation of $b_2^{\perp} = b_2^{\parallel} = b_2^{\parallel\perp}$ from \sctn{sctn:oneconst}, then the deformation free energy can be written in terms of the layer displacement field $u=\phi/q_0$ to be
\begin{subequations}
\begin{align}
   f^\text{comp}+f^\text{curv} &= \frac{B}{2}\left(\nabla_\parallel u\right)^2 + \frac{K}{2} \left(\nabla^2 u\right)^2 , 
   \label{eq:oneConst_simple}
\end{align}
where $B \equiv \varrho q_0^2 \ps^2b_1^\parallel$ and $K \equiv \varrho q_0^2 \ps^2 b_2^{\parallel\perp}$. 
Although not the most widespread model, this form finds use in research on the behaviour of smectic systems. For example, it has recently been used to study screw dislocations~\cite{Aharoni2017} and colloidal inclusions~\cite{gharbi2018}. 

\paragraph{Perpendicular superiority}
Far more common than the one-constant approximation is the assumption that one elastic constant dominates over the others. 
Unlike the one-constant approximation of \eq{eq:oneConst_b2} or \eqref{eq:oneConst_simple}, this second assumption gives precedence to the in-plane deformations representing the bending of layers, as in membrane elastic theory~\cite{Helfrich1994}. 
If we assume that only the $b_2^\perp$ terms contribute to the curvature free energies, while the others are zero, $b_2^{\parallel} = b_2^{\parallel\perp} = 0$, the free energy density reduces to  
\begin{align}
   f^\text{comp}+f^\text{curv} &= \frac{B}{2}\left(\nabla_\parallel u\right)^2 + \frac{K'}{2} \left(\nabla_\perp^2 u\right)^2 , 
\end{align}
\end{subequations}
where $K' \equiv \varrho q_0^2 \ps^2 b_2^\perp$. 
Under this assumption, the second term is quadratic in the mean curvature, $\nabla_\perp^2 u$\cite{alageshan2017}.
This is the the linearized form of the smectic deformation free energy~\cite{Lubensky1997, chaikin1995, Zhang2012, Zhang2013, ferreiro2018}, which is sometimes referred to as the de~Gennes--McMillan form~\cite{mcmillan1971, degennes1972} and other times called the more general Landau--Peierls free energy~\cite{prost1984, chaikin1995, alageshan2017, Zhai2021}. 

The description given above does not account for the microscopic density variations of the density wave. 
If covariant derivatives had been employed, the results are ultimately the same but with \eq{eq:covar} leading to
\begin{align}
    \frac{f^\text{comp}}{\varrho\ps^2} &= b_1^\perp\left( \del\phi\cdot\del\phi  -2q_0\vec{N}\cdot\del\phi +q_0^2\right) 
    \nn \\ &\qquad
    + \left(b_1^\parallel - b_1^\perp\right)\left[ (\vec{N}\cdot\del\phi)^2  -2q_0\vec{N}\cdot\del\phi +q_0^2 \right]^2 . 
\end{align}
If the one constant approximation is made for $b_1$, the second term is eliminated and this reduces to
\begin{align}
    f^\text{comp} &= B'\left(\abs{\del\phi}-q_0\right)^2,
\end{align}
with $B'\equiv \varrho q_0^2 \ps^2b_2^\perp$. 
Free energies of this form are used extensively to study elasticity effects in microscale smectic systems~\cite{alexander2012, Kamien2016, Machon2019, Zappone2020}. 
Furthermore, if we again write this in terms of the layer displacement $u=\phi/q_0$, then the free energy is 
\begin{align}
    f^\text{comp} &= B''\left(\abs{\del u}-1\right)^2,
\end{align}
for $B''\equiv B' q_0^2$, which is another pervasive form for smectic models~\cite{weinan1997}.

\subsubsection{Oseen constraint}
Smectics theories commonly assume twist to be prohibited as a consequence of near incompressibility of the layers. 
This is referred to as the {\em Oseen constraint}~\cite{stewart2007}, which has been employed extensively in smectic liquid crystal theories. 
The prohibition against twist can be understood by noting that the twist $\vec{N}\cdot\left[\del\times\vec{N}\right]$ and bend $\vec{N}\times\left[\del\times\vec{N}\right]$ are both directly proportional to the curl $\del\times\vec{N}$. 
However, as stated at the end of \sctn{sctn:layers}, in traditional theories the normal is entirely determined by phase gradients as $\vec{N} = \del\phi/\abs{\del\phi}$. 
Therefore, the twist (and bend) are proportional to $\del\times\vec{N} \sim \del\times\left(\del\phi\right)$ but the curl of a gradient of a well-behaved scalar field is always zero, which means that twist and bend must be prohibited in this limit~\cite{stewart2007,weinan1997}. 
On the other hand,  bending of the layers themselves (which is equivalent to {\em splaying} of the layer normal) is effortless. 
This is clear from the existence of focal conic domains, which are layers with significant layer-bend but little layer compression~\cite{kim2009a,kim2009b,kim2010a,kim2010b}. 

If we insist that the layers are highly incompressible, the possible twist and bend distortion modes of the layer normal should come at a high free energy cost, while splay of the layer normal (bending of the layers themselves) should not. 
Since our goal is to consider how our model relates to the three modes of nematic distortion, we consider only the first order derivatives in $f^\text{comp}$ and the neglect higher order terms that are in $f^\text{curv}$. 
Under the conditions of fixed $\psi$ but variable $\vec{N}$, the first order derivatives are
\begin{align}
    E_{ij,k}E_{ij,l}^* &= 2\ps^2\nabla_k N_i\nabla_l N_j .
    % \\E_{ij,kk}E_{ij,ll}^* &= 2\ps^2\left(2\nabla_k N_i\nabla_k N_j\nabla_l N_i\nabla_k l_j +\nabla^2 N_i \nabla^2 N_j +N_i N_j \nabla^2 N_i \nabla^2 N_j\right).  
\end{align}
Substituting these into \eq{eq:nondim}, we find 
\begin{align}
    \label{eq:def}
    \frac{f^\text{comp}}{2\ps^2} 
    % &= b_1^\perp\nabla_i N_j \nabla_i N_j + \left(b_1^\parallel-b_1^\perp\right)N_i N_j \nabla_i N_k \nabla_j N_k \\
    &= b_1^\perp \left(\left(\vec{N}\cdot\left[\del\times\vec{N}\right]\right)^2 + A_4\right)+ b_1^\parallel \left(\vec{N}\times\left[\del\times\vec{N}\right]\right)^2. 
\end{align}
The first term is proportional to the twist $\vec{N}\cdot\left[\del\times\vec{N}\right]$, while the second term $A_4=\nabla_i N_j \nabla_j N_i$ can be omitted through an application of Gauss's theorem, which transforms $A_4$ into a surface term~\cite{berreman1984} (though this may have consequences for frustrated structures that cannot fill space such as twist-grain boundaries~\cite{selinger2021} and may limit allowed configurations since it is associated with the Gaussian curvature~\cite{Liarte2015}). 
The third term is proportional to the bend $\vec{N}\times\left[\del\times\vec{N}\right]$. 
High $b_1^\perp$ prohibits twist deformations and $b_1^\parallel$ can be increased to prohibit bend of the layer normal. 
Splay deformations proportional to $\del\cdot\vec{N}$ do not appear in \eq{eq:def} and so splay deformations of the layer normal induce no free energy cost in the incompressible limit.

%-------------------------------------------------------------------%%
%%-------------------------------------------------------------------%%
%%-------------------------------------------------------------------%%
%% Conclusion %%
%%-------------------------------------------------------------------%%
%%-------------------------------------------------------------------%%
%%-------------------------------------------------------------------%%
\section{Conclusion}
This manuscript has presented a phenomenological Landau theory for a complex-tensor order parameter $\tens{E}$.
After introducing smectics and discussing the challenges faced by the complex scalar order parameter description, the success of $\tens{Q}$-tensor theory in nematics to avoid analogous pitfalls in the vicinity of defects was considered. 
This then led us to propose $\tens{E}$ as an order parameter for smectics.
Our $\tens{E}$-tensor formalism encompasses the advantages that $\tens{Q}$-tensor theory provides to nematics but for smectics. 
The tensor is capable of describing the local degree of lamellar ordering, layer displacement, and orientation of the layers. 
It can be described as tensorial, complex, traceless, normal, globally gauge invariant and symmetric but non-Hermitian. 
It also resolves many of the potential ambiguities inherent to complex scalar order parameter models, in a manner that is  mathematically elegant, yet numerically pragmatic since defects can possess a finite core size rather than be point-singularities. 

A phenomenological Landau theory for $\tens{E}$ was created that includes the bulk, compression and curvature free energy. 
The compression contribution results from first order spatial derivatives of the order parameter and has a pair of elastic constant for deformations projected normal to the smectic layers and in-plane deformations. 
Separating these different contribution is made numerically possible through $\tens{E}$-based projection operators. 
Similarly, the curvature term possesses three elastic constants. 
By non-dimensionalizing the free energy, these are seen to correspond to a total of five characteristic length scales. 
The non-perturbed spacing between layers (or equivalently, equilibrium wave number) constitutes a sixth length scale but this is a microscopic scale, while this $\tens{E}$-based description of lamellar materials is appropriate for the hydrodynamic scale. 
Crucially, this model reduces in various limits to currently employed models of simple smectics. 

We hope this work opens new possibilities for numerical studies on smectics possessing many defects, within complex geometries and under extreme confinement. 
While we have attempted to present the mathematical framework of this theory in full, much theoretical work remains to be done. 
At present, this theory is restricted to describing the lamellar/layering properties of liquid crystals alone, rather than the complete smectic phase. 
The orientational properties of nematogens are fully and capably described by the nematic $\tens{Q}$-tensor --- the focus of this study is to adequately express the properties that are purely lamellar through $\tens{E}$, and not nematic in nature. 
True {\em sm-A} or {\em sm-C} liquid crystals will require coupling $\tens{E}$ directly to $\tens{Q}$~\cite{Biscari2007} and this will form a basis for extensions of this model. 
Additionally, we believe that, through its capacity to numerically describe defects, $\tens{E}$-theory will be a powerful addition to smectohydrodynamic descriptions, studies of the rheology of lamellar systems and explorations of intrinsically non-equilibrium materials~\cite{adhyapak2013}.

%%-------------------------------------------------------------------%%
%%-------------------------------------------------------------------%%
%%-------------------------------------------------------------------%%
%% Acknowledgements %%
%%-------------------------------------------------------------------%%
%%-------------------------------------------------------------------%%
%%-------------------------------------------------------------------%%
\section*{Acknowledgements}
TNS thanks the editors for the invitation to contribute to this special issue. 
This research has received funding (TNS) from the European Research Council (ERC) under the European Union’s Horizon 2020 research and innovation programme (Grant agreement No. 851196). 
JP gratefully acknowledges funding from EPSRC. 

%%-------------------------------------------------------------------%%
%%-------------------------------------------------------------------%%
%%-------------------------------------------------------------------%%
%% Movies %%
%%-------------------------------------------------------------------%%
%%-------------------------------------------------------------------%%
%%-------------------------------------------------------------------%%
\section*{Supplementary Movies}
Numerical simulation of perturbed layers in a periodic domain using the one constant approximation. 
        Here, $\Delta x =\xi$ and $\Delta t =0.001\mu$. 
        The physical parameters are $A=-1$, $C=2$ and $\kappa^2=0.6$. 
        We observe a relaxation towards a more ordered state over a total time of $5\mu$. Snapshots from these three movies are shown in \fig{fig:sim}.
        \begin{itemize}
            \item \movie{1} Relaxation dynamics of the phase field $\phi\left(\vec{r},t\right)$. Over time gradients of $\phi$ fade, demonstrating a reduction in layer compression/dilation. 
            The red bars correspond to the layer normal, $\vec{N}\left(\vec{r},t\right)$. 
            \item \movie{2} The extent of the layering $\ps\left(\vec{r},t\right)$ for the same system as in \movie{1}. 
            At later times $\ps$ relaxes to the equilibrium value ($\ps_{eq}=1$).
            \item \movie{3} Same as \movies{1-2} showing the layer visualisation $\RE{\Psi}$ for $q_0=0.913$. 
        \end{itemize}

%%-------------------------------------------------------------------%%
%%-------------------------------------------------------------------%%
%%-------------------------------------------------------------------%%
%% References %%
%%-------------------------------------------------------------------%%
%%-------------------------------------------------------------------%%
%%-------------------------------------------------------------------%%
\section*{References}
\bibliography{refSmectics}

\end{document}